\documentclass[11pt,aps,prd,eqsecnum,superscriptaddress,a4paper]{revtex4-2}
\usepackage{amssymb,amsmath,amsthm,graphicx,amscd}
\usepackage{enumerate,comment,ulem,bm,enumitem}
\usepackage[mathscr]{eucal}
\usepackage[cal=boondoxo]{mathalpha}

\usepackage{xcolor}
\usepackage{tikz}
\usetikzlibrary{arrows.meta}

\usepackage[hidelinks]{hyperref}
\usepackage{orcidlink}

\makeatletter
\renewcommand*\env@matrix[1][\arraystretch]{%
  \edef\arraystretch{#1}%
  \hskip -\arraycolsep
  \let\@ifnextchar\new@ifnextchar
  \array{*\c@MaxMatrixCols c}}
\makeatother

\begin{document}

\title{Vacuum viscosity and relativistic inertia:  Motion of a massive object with charged internal degrees of freedom interacting with a classical field}
\author{Jen-Tsung Hsiang\orcidlink{0000-0002-9801-208X}}
\email{cosmology@gmail.com}
\affiliation{College of Electrical Engineering and Computer Science, National Taiwan University of Science and Technology, Taipei City, Taiwan 106, R.O.C.}
\author{Bei-Lok Hu\orcidlink{0000-0003-2489-9914}}
\email{blhu@umd.edu}
\affiliation{Maryland Center for Fundamental Physics and Joint Quantum Institute,  University of Maryland, College Park, Maryland 20742, USA}
\date{First Version August 8, 2026}

\begin{abstract}
Our present investigation into a rather rudimentary problem is motivated by two classes of problems studied since the 70's, cosmological particle creation and its more accessible analog, the dynamical Casimir effect on the one hand, and quantum friction a neutral atom moving along a dielectric surface would experience, on the other. The backreaction effects of produced particles being able to isotropize the expansion of the universe, or to slow down the moving mirror can be understood via the concept of vacuum viscosity arising from fluctuations of the quantum field. We want to track down the origin of this effect by asking the question whether a moving massive $M$ object with a charged internal degrees of freedom $\chi$  interacting with a free unbounded classical field $\phi$ at zero temperature would experience a viscous force, similar to the said precedents.  Adopting a microphysics model for optomechanics which can treat the unequal tripartite $\chi$-$\phi$-$M$ interactions, we first perform a nonrelativistic calculation, which seems perfectly legitimate considering the needs of atomic physics, and found the answer to be yes, but a relativistic covariant calculation says no. We identify where the nonrelativistic framework is defective. The resolution of this latent yet real conflict is technically nontrivial but physically quite inspirational. It results in added enriched contents to Newton's first and second laws when the principles of special relativity are enforced, and rules to follow to get the correct nonrelativistic answer.       
\end{abstract}
\maketitle

\newpage
\tableofcontents

\section{Introduction}

\subsection{Motive and Aim}

In this paper we aim at finding out whether a massive moving object with internal degrees of freedom carrying a charge, such as a neutral atom, or an Unruh-DeWitt detector \cite{Unr76,DeW79} with mass,  would experience a viscous force from its interaction with a classical field, which we call vacuum viscosity, ``vacuum" here in the sense of zero temperature, free (non-interacting, non-zero amplitude) field in a space with no boundary. What motivated us to ask such a question comes primarily from analyzing the physics of two classes of processes: i) \textbf{Vacuum viscosity}: the  fluctuations of a quantum field when parametrically amplified by the expansion of the universe or, more mundanely,  the motion of a mirror,  give rise to particle creation from the vacuum. Cosmological particle creation \cite{Par69,ZelSta71} and its backreaction on the dynamics of spacetime was explored in the 70's-80's \cite{HuPar77,HuPar78,FisHarHu79,HarHu79,HarHu80,And83}. Similar studies of a moving mirror in a quantum field, as in the dynamical Casimir effect \cite{Dodonov} also commenced in the 70s~\cite{Moore,DavFul77}. When the backreaction of particles production on the dynamical spacetime or the moving mirror is taken into account, vacuum viscosity seems to be responsible for the isotropization of an anisotropically expanding universe \cite{CalHu87,HuSin95} or the slowing down of a moving mirror \cite{WeLee05,XBH,Butera26}. ii) \textbf{Quantum friction}, experienced by a neutral atom moving at uniform speed near a dielectric surface, or even near a conducting plate, a subject which has developed into a subfield in the last three decades. (See some representative papers \cite{Pendry,Fran,Lombardo,Guo23,Farias,Pereira} and informative reviews \cite{Kardar,Reiche}).  

We mention two noteworthy points here: a) \textit{Backreaction nonMarkovian}:  this vacuum viscous force is reactive in nature, like the reactance of an inductor {added to a resistor, not purely resistive} (no phase lag), as “friction” may conjure.  Whenever the backreaction of what is produced in the field,  acting as an environment from an open system viewpoint, is included in the consideration, the dissipative dynamics of the system, here, the spacetime dynamics or moving mirror, is intrinsically non-Markovian, {referring to time-nonlocal, with memory}.  b) \textit{Tripartite interplay}: in the cosmological particle creation and the dynamical Casimir effect cases, only two parties are involved, spacetime dynamics or moving mirror (assumed fully reflective), and the quantum field, whereas in the case of quantum friction experienced by a moving atom, or an imperfect mirror, the internal degrees of freedom (idf), e.g., the electronic activities of the atom or the imperfect mirror, are the party which interacts with the quantum field, not the center of mass of the atom or the mirror, which we may refer to as the external or mechanical degree of freedom (mdf), but has an indirect effect on its motion, which is the focus of our present study. The interplay between the idf and the mdf through direct and indirect interactions with the field in this tri-partite system is a lot more complicated than the bi-partite examples mentioned earlier. (A celebrated case is Sisyphus cooling of an atom \cite{Nobel97}, but for our purpose here, we do not want any control laser to intervene)  We shall describe further our modeling below.

\subsection{Vacuum viscosity: classical vs quantum}

\noindent \textbf{Vacuum viscosity:} The conception and investigation of vacuum viscosity predated modern quantum friction discovery: It was first proposed in 1969 by Zel'dovich \cite{Zel70} to describe the backreaction effects of vacuum particle production on the dynamics of the early Universe. This idea was quantified and made precise by the rigorous work of many authors in the 70’s and 80's mentioned above, using quantum field theory in curved spacetime \cite{BirDav82,ParTom09} methodologies and semiclassical gravity theory \cite{HuVer} for treating backreaction effects \cite{Wald,Thiemann}, captured in terms of vacuum viscosity \cite{HuVacVis}. The one-to-one correspondence between cosmological particle creation and the dynamical Casimir effect \cite{CPC-DCE} enables one to understand the backaction effects of dynamically excited vacuum fluctuations in terms of a quantum Lenz law \cite{HuGraEnt,Par83,XBH}.

\noindent\textbf{Our modest goal:} Both vacuum viscosity and quantum friction involve quantum fields. We shall treat quantum fields in a sequel paper. Here we assume a classical field, so we know it is not due to quantum vacuum fluctuations and particle creation. We consider an atom with  internal degrees of freedom which carry charges (better qualified below) in motion like what is done in many studies of quantum friction, but we do not want the presence of any boundary, no conducting plates nor dielectric medium surfaces. Also, we want to exclude thermal effects, thus we are looking for possible effects of vacuum viscosity due to a classical field at \textit{zero temperature}, which rules out the Einstein-Hopf thermal drag \cite{EinHop}. Finally, the viscosity we look for is distinct from classical radiation reaction.  If we name the trajectory of the atom to be $\bm{z}(t)$, radiation reaction would be from the $\dddot{\bm{z}}(t)$ or 5th order time derivative $z^{(5)}$ terms, whereas we are interested in viscous effects stemming from the $\dot{\bm{z}}(t)$. (The $\ddot{\bm{z}}(t)$ term can be grouped into a time-dependent renormalized effective mass.)

\subsection{Newton's first and second law: nonrelativistic versus relativistic framework}

This classical minimalist set-up naturally brings us to think about the foundations of classical mechanics:  Newton's first law of inertia -- its domain of validity, and Newton's second law -- the notion of mass in the equation of motion.   One can even ask, if vacuum viscosity exists in the setting we described above, an atom expected to move with uniform speed would slow down -- wouldn't that situation contradict Newton's first law? To bring Newton's laws to the context of our present issue it would be fair to upgrade them to modern physics, specifically, incorporating its two pillars: special relativity and quantum mechanics.  We shall do the former here and leave the latter to a later paper.  Focusing on special relativity principles alone, we shall first carry out a calculation in a nonrelativistic framework, followed by another one in a fully relativistic, covariant framework. The differences are quite telling: the  nonrelativistic treatment says Yes, and  the relativistic treatment says No. A particle physicist/field theorist would insist that only a fully relativistic treatment is trustworthy, it is the only way to produce the correct result. {In fact, those with a keen eye after reading the title may come up with an answer, which we certify after some effort,  is the correct one, by invoking Lorentz invariance. They would argue that the physics involving an object moving at uniform speed is no different from it being at rest, then add on the observation that most objects are composites anyway, meaning they possess internal degrees of freedom.  We know this argument is perhaps the most economical, even elegant, however, we do not want to invoke symmetry or invariance principles ab initio, but to see them coming out. We prefer to analyze how each of the three parties involved acts, starting with the interaction between the idf and the field and see how that affects the motion of the object. What we are doing here is like  reverse engineering, or looking under the hood, knowing the whole is beautiful but wanting to see how every part works to make the whole so simple and beautiful. In practice this is not unlike how nuclear and particle theorists examining Lorentz invariance in the internal structures of the nucleons, in the different sectors of their theories, e.g., \cite{Finkel,Nambu,Ji,Liberati} or the design  of experiments to test the magnitudes of Lorentz invariance violation (LIV) in different circumstances and set-ups, e.g, \cite{ColGla,Kost,Chu,Muller}. We will continue this discussion in the Discussion section.}

All this nicely said, and granted, but, we ask,  how are we going to explain to our atomic physicist colleagues who maintain that the atoms in their experiments move slowly, never ever close to the speed of light?  The underlying reasons are quite educational, as we shall see.  Suffice it to say here that the crucial reason why the $\dot{\bm{z}}(\tau)$ term does not manifest in a relativistic treatment is due to the mass-energy relation in special relativity: they are  interchangeable.  The energy in the internal degrees of freedom can be incorporated in the mass renormalization.  This has two immediate consequences, a) in a fully relativistic description, such an object  would follow inertial motion unaltered, Newton's first law holds;  b) the concept of a time-dependent \textit{relativistic inertia} which need be admitted to the application of Newton's second law.  These are the main takeaways from our investigation into a basic issue in classical physics with a bare minimalist set-up and a functional microphysics modeling.

\subsection{A microphysics model for tripartite interplay}\label{S:model}

\subsubsection{The AMOF model} 

The MOF model \cite{MOF1,MOF2,MOF3,MOF4,MOF5,MOF6} was proposed as a theory of quantum optomechanics, for treating the interaction of an imperfect mirror with a quantum field. Traditional treatment is by imposing the suitable boundary conditions on the field at the position of the mirror. The MOF model introduces a microscopic variable, the `mirror oscillator' to describe the optical properties of an imperfect mirror,  thus removing the need to stipulate boundary conditions. Replacing the mirror by an atom, the MOF model captures their electronic activities interacting with the external field, similar to the single oscillator model for describing the dynamical polarizability of an atom \cite{Babb}. One could thus call it the AMOF model (for Atom/Mirror Oscillator model \cite{MOF3}. The interplay between these three dynamical variable can capture the qualitative behaviors of real atoms and mirrors reasonably well \cite{MOF6}.

\subsubsection{The three interlinked dynamical variables}

We adopt  this AMOF microphysics model  to describe the dynamics of a massive object with internal structure. It comprises three interlinked dynamical variables i)  its center-of-mass $M$ in a nonrelativistic treatment, which we refer to as the mechanical (or external) degree of freedom (mdf),  moving along a spacetime trajectory $\bm{z}$ and ii) an internal degree of freedom (idf)  $\chi$, modeled here as a harmonic oscillator with mass $m$ and  charge $\lambda$, interacting with iii) an ambient massless scalar field $\phi$. 
Since the system is treated as a pointlike entity, $\chi$ and $\phi$ are bilinearly coupled  along the object's worldline, meaning, the local field dynamics that drives the internal dynamics are evaluated exactly at $\bm{z}$.
This framework is an effective model for a neutral atom: we can call it a two-level atom if two-level activities is the focus of attention, or a harmonic atom if the internal space variable $\chi$ undergoes harmonic motion,  representing an induced, non-permanent dipole. 

For simplicity we work with a scalar field  here coupled with a monopole charge representing the internal dof. In the real world situation what this model emulates would be for electrons in an oscillating dipole distribution interacting with an external electromagnetic field in a derivative type of coupling.  The difference from a full treatment of a real neutral atom is our ignoring the  positive charge carried by the nucleus in balance with that of the electron. Since our focus here is on how the idf interacts with an external field and how that affects the motion of the mechanical dof, the mass $M$, we shall leave out the  interaction between the charged mass $M$ with the field with little effect on the issue at hand~\footnote{While successful for explaining the essential physics of stationary or inertial atoms, this model does not work well for arbitrary trajectories. In a physical charge-neutral atom, the massive nucleus carries a charge opposite to the electron cloud and will inevitably generate its own radiation during acceleration. This introduces complex shielding and interference effects with the surrounding internal charge distribution that the model ignores. Furthermore, because the net charge in this effective setup may not be identically zero, it cannot produce the exact radiation profile of a strictly neutral atom. In practical terms, this model is probably better suited for describing time-dependent macroscopic charge distributions, microscopic dipolar emitters, or moving mirrors.}. In physical terms, this setup is more like a small antenna (point-like) with mass $M$, and oscillating electrons in a dipole charge distribution (not a dipole formed by two opposite charges). 

Another useful representation is an Unruh-DeWitt (UdW) detector \cite{Unr76,DeW79,MOF3} which is an idealized entity used for the exploration of acceleration radiation. Usually one assumes that the UdW detector is point-like and massless, with a two-level  or a harmonic oscillator configuration for its idf.  Our model depicts an UdW detector with mass \cite{SudKem}.

We give a quick description of how to find out what we want from the way how the three dynamical variables are linked to each other:  There is direct interaction between the idf and the field. In a nonrelativistic treatment there is no direct interaction between the idf and the mdf.  There is linkage, not interaction,  between the mdf and the field,  in the sense that the field takes on  values at the spacetime point where the moving atom with mass $M$ is located. However, in a relativistic treatment where the coordinate (lab) time $t$ is replaced by the proper time $\tau$, the  the velocity of the mdf enters in the dynamics of the idf through the gamma factor. For both cases, the primary challenge is to see how the idf-field interaction affects the motion of the mdf.

{This paper is organized as follows. In Sec.~\ref{S:epgff}, we present a nonrelativistic treatment of the tripartite system to illustrate how treating the mechanical and internal dynamics nonrelativistically while coupling them to a relativistic field leads to spurious kinematic artifacts, which could mislead one to the belief of the presence of vacuum viscosity. To ensure total theoretical integrity  Sec.~\ref{S:engnge} establishes a fully relativistic covariant formulation for the scalar field and the internal degrees of freedom. A covariant approach is necessary to correctly account for mass-energy equivalence and to prevent  artifacts of coordinate origins from masquerading as genuine physical backreaction effects. Section~\ref{S:otgyor} then extends this covariant framework to the mechanical degree of freedom, addressing the emergence of a dynamic, proper-time dependent mass and the requisite renormalization procedures. Finally, Sec.~\ref{S:obyrw} summarizes our key findings, discusses several conceptual issues for future extensions involving quantum fields, and ends with a few take-home messages. Supporting mathematical derivations are provided in Appendixes~\ref{S:eouhegd}--\ref{S:eothgbd}.}

\section{Nonrelativistic treatment}\label{S:epgff}

In a nonrelativistic treatment, one uses the coordinate time $t$ in the lab frame. The total action of the tripartite interacting system described in Section~\ref{S:model} is the sum of the actions of each subsystem plus the interaction term, $S=S_{\text{mdf}}+S_{\text{idf}}+S_{\text{field}}+S_{\text{int}}$. We begin with the action for the mechanical (external) degree of freedom $S_{\text{mdf}}[\bm{z},\dot{\bm{z}}]$ is 
\begin{equation}
	S_{\text{mdf}}[\bm{z},\dot{\bm{z}}]=\int\!dt\;\biggl[\frac{M}{2}\dot{\bm{z}}^{2}-V(\bm{z})\biggr]\,,
\end{equation}
where $M$ is the mass of the atom (or an Unruh-DeWitt detector with mass) and $V(\bm{z})$ is some external potential. In this section, an overdot indicates taking the derivative with respect to time $t$ in the lab frame, as is customary in the description of nonrelativistic motion. The action for the internal degree of freedom $S_{\text{idf}}[\chi,\dot{\chi}]$ takes the form 
\begin{equation}
	S_{\text{idf}}[\chi,\dot{\chi}]=\int\!dt\;\biggl[\frac{m}{2}\dot{\chi}^{2}-\frac{m\omega^{2}}{2}\chi^{2}\biggr]\,,
\end{equation}
with $m$ being the mass of the internal degree of freedom and $\omega$ identified as the bare frequency or energy spacing.  The action for the free field $\phi(t,\bm{x})$ is given by 
\begin{equation}
	S_{\text{field}}[\phi,\dot{\phi}]=\int\!d^{4}x\;\biggl[\frac{1}{2}\bigl(\partial_{t}\phi\bigr)^{2}-\frac{1}{2}\bigl(\bm{\nabla}_{x}\phi\bigr)^{2}\biggr]\,,
\end{equation}
where the spacetime point  $x^{\mu}=(t,\bm{x})$ is denoted by $x$. Finally, the interaction action $S_{\text{int}}[\bm{z},\chi,\phi]$ is 
\begin{equation}
	S_{\text{int}}[\bm{z},\chi,\phi]=\int\!d^{4}x\;\biggl[\lambda\,\chi(t)\phi(t,\bm{x})\,\delta^{(3)}(\bm{x}-\bm{z}(t))\biggr]\,.
\end{equation}
where $\lambda$ is the coupling constant between the idf and the field. (It {represents the scalar charge $q$} if coupled to a scalar field, or the electric charge $e$ if coupled to an electromagnetic field, in which case one needs to introduce two polarizations for a vector field. )

The corresponding equations of motion are given by
\begin{align}
	\bigl(\partial_{t}^{2}-\bm{\nabla}_{\bm{x}}^{2}\bigr)\phi(t,\bm{x})&=\lambda\,\chi(t)\,\delta^{(3)}(\bm{x}-\bm{z}(t))\,,\label{E:oehdgd}\\
	m\,\ddot{\chi}(t)+m\omega^{2}\,\chi(t)&=\lambda\,\phi(t,\bm{z}(t))\,,\label{E:ytwtrye}\\
	M\,\ddot{\bm{z}}(t)+M\Omega^{2}\,\bm{z}(t)&=\lambda\,\chi(t)\,\bm{\nabla}_{\bm{z}}\phi(t,\bm{z}(t))\,.\label{E:oreoti}
\end{align}

\subsection{Field and internal dof dynamics}

Formally solving Eq.~\eqref{E:oehdgd} for $\phi$ gives
\begin{align}\label{E:ndlgld}
	\phi(t,\bm{x})&=\phi_{\text{h}}(t,\bm{x})+\phi_{\text{ret}}(t,\bm{x})\,,&\phi_{\text{ret}}(t,\bm{x})&=\lambda\!\int_{0}^{t}\!dt'\;G_{\textsc{r}}^{(\phi)}(t,\bm{x}\,;\,t',\bm{z}(t'))\,\chi(t')\,,
\end{align}
Here, the second term $\phi_{\text{ret}}$ on the right is the scalar version of the Li\'enard-Wiechert potential, produced by the internal dynamics $\chi(t)$. The first term $\phi_{\text{h}}$ is the pre-existing free scalar field. The two-point function G$_{\textsc{r}}^{(\phi)}(t,\bm{x}\,;\,t',\bm{x}')$ is the retarded Green's function constructed from the free field $\phi_{\text{h}}$, not the full field $\phi$. The back-action of this retarded field will contribute to the frequency renormalization and the damping force on internal dynamics, each of which can be clearly identified as the consequences of the bond field $\phi_{\text{bnd}}$ and the radiation field $\phi_{\text{rad}}$.
\begin{align}\label{E:vndof}
    \phi_{\text{rad}}(x)&=\frac{1}{2}\Bigl[\phi_{\text{ret}}(x)-\phi_{\text{adv}}(x)\Bigr]\,,&\phi_{\text{bnd}}(x)&=\frac{1}{2}\Bigl[\phi_{\text{ret}}(x)+\phi_{\text{adv}}(x)\Bigr]\,,
\end{align}
corresponding to the symmetric and the anti-symmetric superpositions of the retarded and advanced fields. By construction, the radiation field satisfies the free-field wave equation $\square \phi_{\text{rad}}(x)=0$, while the bond field obeys $\square \phi_{\text{rad}}(x)=\varrho(x)$ with the source density $\varrho(x)=\lambda\,\chi(t)\,\delta^{(3)}(\bm{x}-\bm{z}(t))$.

These effects can be identified in the nonlocal expression in the reduced equation of motion of the internal degree of freedom after we substitute Eq.~\eqref{E:ndlgld} back to Eq.~\eqref{E:ytwtrye},
\begin{align}\label{E:gort}
	m\,\ddot{\chi}(t)+m\omega^{2}\,\chi(t)-\lambda\,\phi_{\text{ret}}(t,\bm{z}(t))&=\lambda\,\phi_{\text{h}}(t,\bm{z}(t))\,.
\end{align}
The explicit expression of the retarded Green's function $G_{\textsc{r}}^{(\phi)}(x;x')$ of the free field takes the form
\begin{equation}
	G_{\textsc{r}}^{(\phi)}(x;x')=\frac{1}{2\pi}\,\theta(t-t')\,\delta((x-x')^2)\,,
\end{equation}
where $x$ is the shorthand notation of the spacetime point, $x^{\mu}=(t,\bm{x})$. The delta function ensures that the retarded Green's function has support only on the lightcone, while the unit-step function enforces retardation. Thus, the retarded and the advanced fields are given by
\begin{equation}
    \phi_{\substack{\text{ret}\\\text{adv}}}(x)=\frac{1}{4\pi}\int\!d^{3}x'\;\frac{\varrho(t\mp r,\bm{x}')}{r}\,,
\end{equation}
where $r=\lvert\bm{x}-\bm{x}'\rvert$. Eventually, since we will evaluate these fields directly on the detector's worldline, we will substitute $\bm{x}$ with $\bm{z}(t)$ and $\bm{x}'$ with $\bm{z}(t')$ respectively. Since the source density $\varrho$ is highly localized around the detector, the integration variable $\bm{x}'$ is confined to the immediate vicinity of the worldline. Consequently, the distance $r$ is treated as fixed and small, allowing us to Taylor-expand these fields around the current time $t$,
\begin{equation}
    \phi_{\substack{\text{ret}\\\text{adv}}}(x)=\frac{1}{4\pi}\int\!d^{3}x'\;\frac{1}{r}\Bigl\{\varrho(t,\bm{x}')\mp r\,\dot{\varrho}(t,\bm{x}')+\frac{r^{2}}{2}\,\ddot{\varrho}(t,\bm{x}')\mp\frac{r^{3}}{6}\,\dddot{\varrho}(t,\bm{x}')+\cdots\Bigr\}\,,
\end{equation}
so that 
\begin{align}
	\phi_{\text{rad}}(x)&=-\frac{1}{4\pi}\int\!d^{3}x'\;\Bigl\{\dot{\varrho}(t,\bm{x}')+\frac{r^{2}}{6}\,\dddot{\varrho}(t,\bm{x}')+\cdots\Bigr\}\,,\\
    \phi_{\text{bnd}}(x)&=+\frac{1}{4\pi}\int\!d^{3}x'\;\Bigl\{\frac{\varrho(t,\bm{x}')}{r}+\frac{r}{2}\,\ddot{\varrho}(t,\bm{x}')+\cdots\Bigr\}\,.
\end{align}
Eq.~\eqref{E:gort} then becomes
\begin{equation}\label{E:ohwot}
    \ddot{\chi}(t)+\biggl[\omega^{2}-\frac{\lambda^2}{4\pi mr}\biggr]\,\chi(t)+\frac{\lambda^2}{4\pi\,m}\,\dot{\chi}(t)=\frac{\lambda}{m}\,\phi_{\text{h}}(t,\bm{z}(t))\,,
\end{equation}
to the leading order of the expansion. Introduce the renormalized frequency $\omega_{\text{ren}}^2=\omega^2-\lambda^2/(4\pi mr)$. With $r\to 0$,  we arrive at a very simple equation of motion for the internal degree of freedom
\begin{equation}
    \ddot{\chi}(t)+2\bar{\gamma}\,\dot{\chi}(t)+\omega_{\text{ren}}^2\,\chi(t)=\frac{\lambda}{m}\,\phi_{\text{h}}(t,\bm{z}(t))\,,
\end{equation}
where $\bar{\gamma}=\lambda^2/8\pi m$.

This na\"ive expansion seems to satisfactorily reproduce the familiar expression for internal dynamics. However, it is somewhat surprising that the external velocity does not play any role. Tracing back the previous derivation, we note that it is faulty to assume that the separation $r$ is fixed when the detector is in motion, described by a time-dependent variable $\bm{z}(t)$. To carefully take the external motion into consideration, observe that if we let $t'=t-u$, the retarded field in Eq.~\eqref{E:ndlgld} can be cast into
\begin{equation}
	\phi_{\text{ret}}(x)=\frac{\lambda}{2\pi}\int_{0}^{\infty}\!du\;\chi(t-u)\,\theta(u)\,\delta(\sigma-\varepsilon^{2})\,,
\end{equation}
where we have introduced an infinitesimal invariant scalar cutoff $\varepsilon$  to regularize the singularity of the delta function in the retarded Green's function, and extended the upper limit of integration to infinity since the contribution is strictly localized near $u\to0$. Here, $\sigma(u)=(t-t')^2-\lvert\bm{z}(t)-\bm{z}(t')\rvert^2=u^2-\lvert\bm{z}(t)-\bm{z}(t-u)\rvert^2$ is the squared invariant length. Following the derivations in Appendix~\ref{S:eouhegd}, we find that  the divergent part of $\phi_{\text{ret}}$ is given by
\begin{align}
	\phi_{\text{ret}}^{(\text{fr})}&=\frac{\lambda}{4\pi\varepsilon}\,\gamma_{\bm{v}}(t)\,\chi(t)\,,
\end{align}
where $\gamma_{\bm{v}}=(1-\bm{v}^2)^{-\frac{1}{2}}$ is the Lorentz factor, while the finite part becomes
\begin{align}\label{E:riiikf}
	\phi_{\text{ret}}^{(\text{finite})}&\simeq-\frac{\lambda}{4\pi}\,\bigl[1+\bm{v}^{2}(t)\bigr]\,\dot{\chi}(t)-\frac{\lambda}{4\pi}\,\bigl[\bm{v}(t)\cdot\bm{a}(t)\bigr]\,\chi(t)\,.
\end{align}
to the leading non-trivial order in the nonrelativistic limit $\lvert\bm{v}\rvert\ll1$.

We obtain an improved version of Eq.~\eqref{E:gort} as
\begin{equation}\label{E:dbskgs}
    \ddot{\chi}(t)+\biggl[\omega^{2}-\frac{\lambda^2}{4\pi\,m\,\varepsilon}+\frac{\lambda^2}{4\pi}\,\bigl(\bm{v}\cdot\bm{a}\bigr)\biggr]\,\chi(t)+\frac{\lambda^2}{4\pi\,m}\,\bigl(1+\bm{v}^{2}\bigr)\,\dot{\chi}(t)=\frac{\lambda}{m}\,\phi_{\text{h}}(t,\bm{z}(t))\,.
\end{equation}
 {Thus, a more careful treatment leads to two additional contributions, which are most transparent in the retarded field given in Eq.~\eqref{E:riiikf}. The first term on the right hand side of Eq.~\eqref{E:riiikf} is odd under time reversal and therefore represents the familiar dissipative force acting on the internal oscillator. This dissipative contribution is amplified by $1+\bm{v}^{2}$, indicating that, to this order, the internal damping is enhanced by the  motion of the whole system. Physically, relativistic time dilation and the distortion of the surrounding field geometry force the moving oscillator to radiate energy into the scalar field at a higher rate than it would at rest.}

 {The second term on the right-hand side of Eq.~\eqref{E:riiikf} does not represent damping. Its coefficient is proportional to the total time derivative of $\bm{v}^{2}$ and is therefore not positive definite. More specifically, it originates from the time derivative of the Lorentz factor, $\dot{\gamma}_{\bm{v}}=\gamma_{\bm{v}}^{3}\,(\bm{v}\cdot\bm{a})$. This contribution is nonzero when the atom undergoes longitudinal acceleration, so that the magnitude of its velocity changes with time. Rather than damping the system, this non-inertial effect dynamically corrects the oscillation frequency or the effective energy gap of the internal state.}

 {The resulting equation of motion, Eq.~\eqref{E:dbskgs}, takes the form of a parametric differential equation. Since these coefficients are continuously modulated by the external variable kinematics $\bm{v}(t)$ and $\bm{a}(t)$, which themselves will be dynamically determined by the coupled equation of motion for $\bm{z}(t)$, the composite system represents a highly nonlinear feedback loop rather than a simple driven oscillator. The internal subsystem cannot conserve its own energy, as this parametric dependence mathematically enforces a continuous, two-way exchange of energy between the external center-of-mass motion and the internal dynamics. In particular, this dynamical feedback introduces a fundamental competition that undermines the mechanical stability of the detector. During the stage of deceleration, $\bm{v}\cdot\bm{a}<0$, the non-inertial correction softens the effective oscillating frequency of $\chi$, seemingly opening the door to parametric instabilities where the effective squared frequency could become negative. However, this apparent mechanical instability is purely an artifact of the coordinate-time parameterization. As we shall see, when evaluated in the fully relativistic formulation using the detector's proper time $\tau$, the equation of motion simplifies to Eq.~\eqref{E:foeue} below, taking the form of a stable damped harmonic oscillator with constant coefficients. The apparently destabilizing parametric modulations in the lab frame are thus revealed to be purely kinematic consequences of projecting proper-time derivatives onto the laboratory coordinate time. Consequently, the fully covariant formulation guarantees the absolute invariant mechanical stability of the internal dynamics, effectively resolving the apparent dynamical instabilities in a nonrelativistic description.}

\subsection{Mechanical (external) dof dynamics} 

\begin{figure}[htbp]
    \centering
    \includegraphics[width=0.5\linewidth]{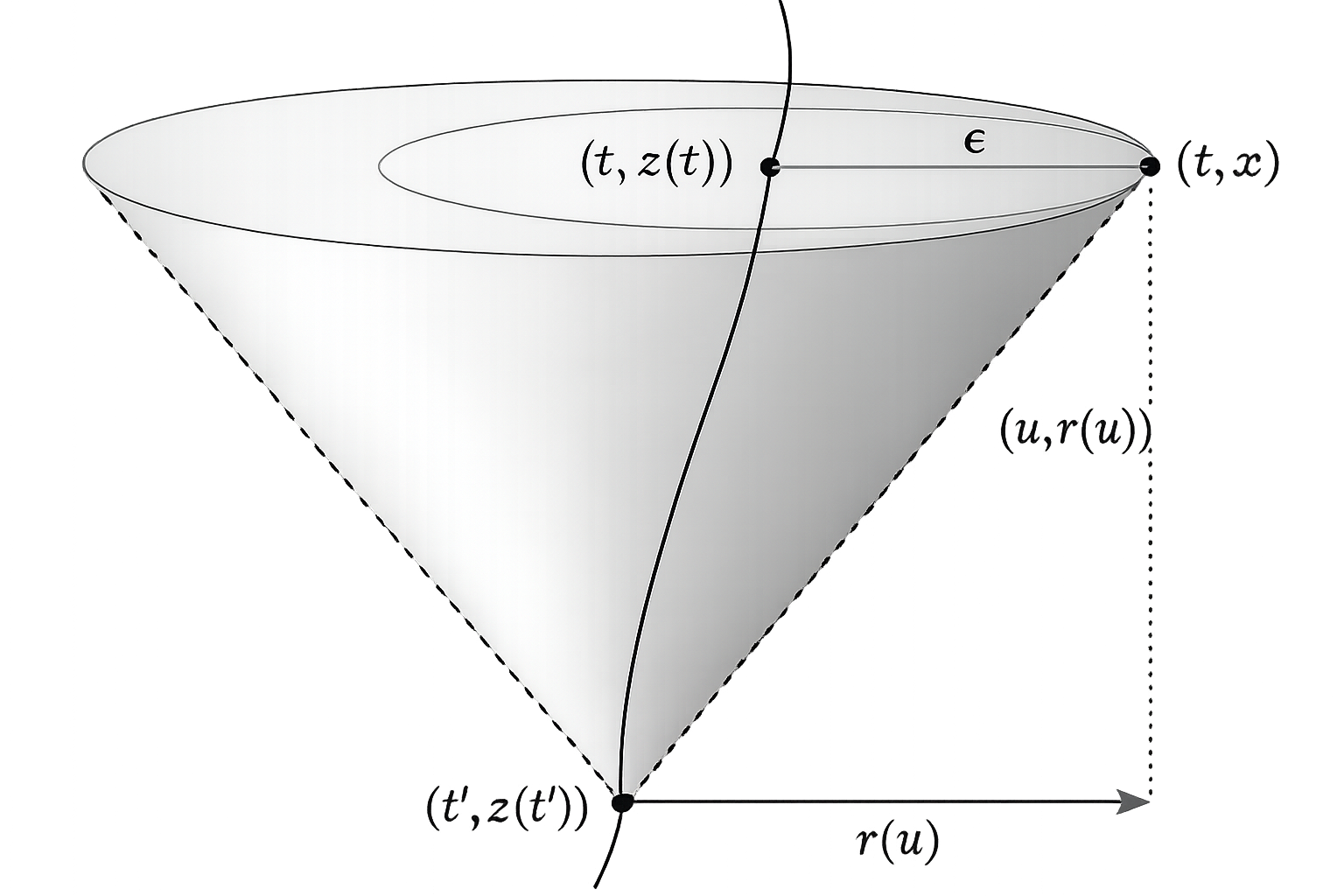}
    \caption{Spacetime diagram illustrating the future lightcone of $\bm{z}(t')$ intersecting a regularizing sphere of radius $\bm{\epsilon}$ centered on the current position $\bm{z}(t)$.}
    \label{Fi:lcReg}
\end{figure}

Next, we reduce the equation of motion for $\bm{z}$, Eq.~\eqref{E:oreoti}
\begin{equation}\label{E:rmtrio}
    M\,\ddot{\bm{z}}(t)+M\Omega^{2}\,\bm{z}(t)-\lambda^{2}\,\chi(t)\int_{0}^{t}\!dt'\;\bm{\nabla}_{\bm{z}}G_{\textsc{r}}^{(\phi)}(t,\bm{z}(t)\,;\,t',\bm{z}(t'))\,\chi(t')=\lambda\,\chi(t)\,\bm{\nabla}_{\bm{z}}\phi_{\text{h}}(t,\bm{z}(t))\,.
\end{equation}
To evaluate the gradient of the nonlocal expression involving the retarded Green's function, we follow Dirac's strategy~\cite{Dirac,Rohrlich} and shift away from the worldline $\bm{z}(t)$ to $\bm{x}$ for the moment by introducing a tiny observation/cutoff distance $\bm{\epsilon}$ (referring to Fig.~\ref{Fi:lcReg} for geometry),
\begin{equation}\label{E:doghoer}
	\lambda^2\chi(t)\int_0^t\!du\;\bm{\nabla}_{\bm{z}} G_{\textsc{r}}^{(\phi)}(t,\bm{x};t',\bm{z}(t'))\,\bigg|_{\bm{x}=\bm{z}(t)+\bm{\epsilon}}\,\chi(t-u)\,.
\end{equation}
where $t'=t-u$. The gradient of the retarded Green's function gives
\begin{align}\label{E:gohort}
	\bm{\nabla}_{\bm{z}} G_{\textsc{r}}^{(\phi)}(t,\bm{x};t',\bm{z}(t'))\,\bigg|_{\bm{x}=\bm{z}(t)+\bm{\epsilon}}&=\frac{1}{2\pi}\,\theta(u)\,\bm{\nabla}_{\bm{z}}\delta(u^{2}-\lvert\bm{r}\rvert^{2})=-\frac{1}{\pi}\,\theta(u)\,\bm{r}(u)\,\delta'(u^{2}-\lvert\bm{r}\rvert^{2})\,,
\end{align}
where the spatial distance vector $\bm{r}(u)=\bm{x}-\bm{z}(t-u)=\bm{\epsilon}+\bigl[\bm{z}(t)-\bm{z}(t-u)\bigr]$ links $\bm{x}$ to the retarded point $\bm{z}(t')$. Then, putting together the results of Eqs.~\eqref{E:dijeor} and \eqref{E:gnhoie} in Appendix~\ref{S:eohodghfg}, we obtain
\begin{align}\label{E:boeif}
	&\quad\lambda^{2}\chi(t)\int_{0}^{t}\!dt'\;\bm{\nabla}_{\bm{z}}G_{\textsc{r}}^{(\phi)}(t,\bm{z}(t)\,;\,t',\bm{z}(t'))\,\chi(t')\notag\\
    &=-\frac{\lambda^2}{6\pi\epsilon}\Bigl(\chi\dot{\chi}\dot{\bm{z}}+\frac{1}{2}\,\chi^{2}\ddot{\bm{z}}\Bigr)+\frac{\lambda^2}{4\pi}\,\Bigl(\chi\ddot{\chi}\dot{\bm{z}}+\chi\dot{\chi}\ddot{\bm{z}}+\frac{1}{3}\,\chi^{2}\dddot{\bm{z}}\Bigr)+\mathcal{O}(\epsilon)\,.
\end{align}
The $\mathcal{O}(1/\epsilon)$ term diverges as $\epsilon\to0$ and a mass renormalization needs be introduced. Note that it can be organized into a total time derivative,
\begin{equation*}
    \frac{d}{dt}\biggl(-\frac{\lambda^2}{12\pi\epsilon}\,\chi^{2}\,\dot{\bm{z}}\biggr)\,.
\end{equation*}
The expression inside the parentheses is quite naturally identified as a divergent momentum correction, $\delta\bm{p}$, where $\bm{p}$ is the momentum conjugate to $\bm{z}$.

On the other hand, the finite $\mathcal{O}(\epsilon^0)$ terms govern the radiation reaction. In the limit where the internal degree of freedom is frozen ($\dot{\chi}=\ddot{\chi}=0$) and we define a constant effective  {dipole $\mathfrak{d}=\lambda\chi$ (which is essentially the scalar analog of the electric dipole)}, this force reduces to the familiar classical scalar Abraham-Lorentz-Dirac force.

We see that when the internal dynamics is included in our consideration, two additional contributions emerge. Taken at face value, the implications of these terms are physically striking: one is proportional to the velocity $\dot{\bm{z}}$, meaning a drag force appears to persist even when the detector is in uniform, unaccelerated motion. If we naively promote these classical results to quantum-mechanical operator expressions, this velocity-dependent dissipation hints at the possibility of quantum vacuum viscosity. Such an effect would cause an inertial detector to spontaneously decelerate, posing a direct challenge to the covariance of the vacuum. This apparent tension calls into question whether this nonrelativistic description correctly captures the full physical picture.

{Note that there is an imbalance in the treatment: the internal and mechanical degrees of freedom are treated non-relativistically while the field dynamics are inherently relativistic.  Hints of discrepancies manifest most clearly in the spontaneous appearance of kinematic terms like the Lorentz factor $\gamma_{\bm{v}}$ and the velocity-acceleration coupling $\bm{v}\cdot\bm{a}$. We need to resolve the apparent conflict in the nonrelativistic description with covariance before deciding whether phenomena like vacuum viscosity are of a genuine physical nature or merely artifacts of an incomplete or restricted treatment. We now proceed to work out the three partite dynamics in a relativistic fully covariant formulation.}

\section{Relativistic covariant dynamics of the field and internal dof}\label{S:engnge}

The covariant dynamics of such an interacting system is described by the action $S=S_{\text{mdf}}+S_{\text{idf}}+S_{\text{field}}+S_{\text{int}}$, where the action of the  mechanical degree of freedom $S_{\text{mdf}}[z]$ is the standard relativistic action for a point particle in an external potential $V$,
\begin{equation}
    S_{\text{mdf}}[z]=-\int\!d\tau\;\bigl[M+V(z(\tau))\bigr]\,,
\end{equation}
where $M$ is the rest mass of the detector \footnote{We switch to this terminology in this section because the Unruh-DeWitt detector has become a popular entity in the gravitation and relativistic quantum information communities \cite{HLL} (even though it was introduced only as an analog to black hole quantum physics).  The focus is on the activities of the internal dof and the treatment is often relativistic, but the effects on the external dof is often neglected.  Here we consider such a detector with mass \cite{SudKem}.} and $z^{\mu}(\tau)$ denotes the detector's worldline parameterized by proper time $\tau$. The action of the internal degree of freedom $S_{\text{idf}}[\chi]$, which is a scalar defined along the worldline, takes the form of a harmonic oscillator, parameterized by $\tau$
\begin{equation}\label{E:ohisgf}
    S_{\text{idf}}[\chi]=\int\!d\tau\;\biggl[\frac{m}{2}\dot{\chi}^{2}-\frac{m\omega^{2}}{2}\chi^{2}\biggr]\,,
\end{equation}
where $m$ is the mass of the internal degree of freedom, $\omega$ is the bare frequency (energy) spacing, and the overdot in this section denotes the derivative with respect to the proper time $\tau$. The action of the free massless scalar field $\phi(x)$ is given in manifestly covariant form by
\begin{equation}
    S_{\text{field}}[\phi]=\frac{1}{2}\int\!d^{4}x\;\partial_{\mu}\phi\partial^{\mu}\phi\,,
\end{equation}
with the spacetime point denoted by $x^{\mu}$. Hereafter, we use the metric convention $(+,-,-,-)$. The interaction action $S_{\text{int}}[z,\chi,\phi]$ couples the internal degree of freedom to the field evaluated on the worldline via a four-dimensional Dirac delta function,
\begin{equation}
    S_{\text{int}}[z,\chi,\phi]=\lambda\int\!d^{4}x\!\int\!d\tau\;\chi(\tau)\phi(x)\,\delta^{(4)}(x-z(\tau)) = \lambda\int\!d\tau\;\chi(\tau)\phi(z(\tau))\,,
\end{equation}
where $\lambda$, which denotes the (scalar) charge of the internal degree of freedom, is a measure of the interaction strength between $\chi$ and $\phi$.  The mechanical degree of freedom appears in the argument of the Dirac delta function, or equivalently, in the argument of the field. Thus, the interaction is still linear in both the internal variable $\chi$ and the scalar field $\phi$, but depends nonlinearly on the mechanical degree of freedom through the field evaluated along the worldline $z^{\mu}(\tau)$, as in the nonrelativistic formulation.

We write down the equations of motion for these dynamical variables by varying the full action $S$ with respect to each of them \text{(Details can be found in Appendix~\ref{S:fbgsfs})}. For the field degree of freedom $\phi(x)$, we have
\begin{align}\label{E:wuhfbds}
    \frac{\delta S}{\delta \phi}&=0\,, &&\implies &\partial_{\mu}\partial^{\mu}\phi(x)&=\lambda\int_{-\infty}^{\infty}\!d\tau\;\chi(\tau)\,\delta^{(4)}(x-z(\tau))\,.
\end{align}
where the right hand side describes a scalar source moving along the worldline of the detector. The equation of motion for the internal degree of freedom $\chi(\tau)$ is
\begin{align}\label{E:vgeor}
    \frac{\delta S}{\delta \chi}&=0\,, &&\implies &m\,\frac{d^2\chi(\tau)}{d\tau^2}+m\omega^{2}\chi(\tau)&=\lambda\,\phi(z(\tau))\,.
\end{align}
As for the mass $M$ carrying a scalar charge,  its mechanical degree of freedom $z^{\mu}$  obeys the equation of motion 
\begin{align}\label{E:voeri}
    \frac{\delta S}{\delta z^{\mu}}&=0\,, &&\implies &\frac{d}{d\tau}\Bigl[M(\tau)\,\frac{d z_{\mu}(\tau)}{d\tau}\Bigr]+\partial_{\mu} V(z(\tau))&=\lambda\,\chi(\tau)\,\partial_{\mu}\phi(z(\tau))\,.
\end{align}
Here, since the mechanical degree of freedom exchanges energy with both the internal oscillator and the scalar field, its inertia is no longer just the bare rest mass $M$. It becomes a time-dependent effective mass
\begin{equation}
    M(\tau)=M+V(z)+\frac{m}{2}\,\biggl(\frac{d\chi}{d\tau}\biggr)^2+\frac{m\omega^2}{2}\,\chi^2-\lambda\,\chi\phi(z)\,.
\end{equation}
We will return to this point in greater detail later. As seen earlier, the equation of motion for $z^{\mu}$ is highly nonlinear due to the scalar interaction term on its right-hand side.

Introducing a fully relativistic description for the internal dynamics may look like an overkill, since the motion of massive objects measurable in the lab is often in the non-relativistic regime, with velocity $\bm{v}=\dot{\bm{z}}(t)\ll1$  small compared to the speed of light. It is tempting to ignore the $\mathcal{O}(\bm{v}^2)$ contribution and simply identify the infinitesimal proper interval $d\tau=\gamma_{\bm{v}}^{-1}\,dt$ with the infinitesimal coordinate time $dt$, where $\gamma_{\bm{v}}=(1-\bm{v}^2)^{-1/2}$ is the Lorentz factor. However, when we take the time derivative of the Lorentz factor $\gamma_{\bm{v}}$, it gives a term proportional to  $\gamma_{\bm{v}}^3\bm{v}\cdot\bm{a}$, which provides an $\mathcal{O}(\bm{v})$ contribution. Hence, if we prematurely drop the $\mathcal{O}(\bm{v}^2)$ contribution at the level of the action, we will completely miss this $\mathcal{O}(\bm{v})$ dynamical contribution for an arbitrarily moving system. This subtlety is an example of why we advocate adhering to the covariant description for the action of the internal degree of freedom, Eq.~\eqref{E:ohisgf}, where the Jacobian of the time integration correctly introduces a factor of $\gamma_{\bm{v}}^{-1}$.

The equations Eqs.~\eqref{E:wuhfbds}, \eqref{E:vgeor} and \eqref{E:voeri} give a simultaneous set of equations for each subsystem
\begin{align}
	   \partial_{\mu}\partial^{\mu}\phi(x)&=\lambda\int_{-\infty}^{\infty}\!d\tau\;\chi(\tau)\,\delta^{(4)}(x-z(\tau))\,,\label{E:fgri}\\
	   m\,\frac{d^2\chi(\tau)}{d\tau^2}+m\omega^{2}\chi(\tau)&=\lambda\,\phi(z(\tau))\,,\label{E:wygrwsf}\\
	   \frac{d}{d\tau}\Bigl[M(\tau)\,\frac{d z_{\mu}(\tau)}{d\tau}\Bigr]+\partial_{\mu} V(\bm{z})&=\lambda\,\chi(\tau)\,\partial_{\mu}\phi(z(\tau))\,.\label{E:wiwiew}
\end{align}
We will then derive the reduced dynamics for the internal and mechanical degrees of freedom. By `reduced' we are using the language of open systems, wherein a  subsystem of the whole, regarded as the environment,  is integrated over to produce an effective dynamics of the subsystem of interest.

\subsection{Scalar field}

Formally solving the field equation for $\phi(x)$, we have
\begin{align}\label{E:dieur}
	\phi(x)&=\phi_{\text{h}}(x)+\lambda\int\!d^{4}x'\!\int_{0}^{\infty}\!d\tau'\;G_{\textsc{r}}^{(\phi)}(x,x')\,\chi(\tau')\,\delta^{(4)}(x'-z(\tau'))\notag\\
	&=\phi_{\text{h}}(x)+\lambda\!\int_{0}^{\infty}\!d\tau'\;G_{\textsc{r}}^{(\phi)}(x,z(\tau'))\,\chi(\tau')\,,
\end{align}
where $\phi_{\text{h}}(x)$ is the homogeneous solution of the field equation and is thus recognized as the free field,
\begin{align}
	\partial_{\mu}\partial^{\mu}\phi_{\text{h}}(x)&=0\,,
\end{align}
and $G_{\textsc{r}}^{(\phi)}(x,x')$ is the retarded Green's function of the free field $\phi_{\text{h}}$, which satisfies the equation
\begin{align}
	\partial_{\mu}\partial^{\mu}G_{\textsc{r}}^{(\phi)}(x,x')&=\delta^{(4)}(x-x')\,.
\end{align}
Here we assume that the interaction is switched on at the proper time $\tau=0$. Thus, the solution Eq.~\eqref{E:dieur} states that after the interaction is switched on, the ambient field in the otherwise empty space is the combination of the pre-existing free field and the retarded field emitted from the internal degree of freedom $\chi(\tau)$. By construction, $\phi(x)$ is determined by the evolutionary history of $\chi(\tau)$.

\subsection{Internal degree of freedom}

Substituting $\phi(x)$ back into the equation of motion for $\chi(\tau)$, Eq.~\eqref{E:wygrwsf}, we obtain the effective equation of motion for the internal degree of freedom,
\begin{equation}\label{E:eituef}
	m\,\frac{d^2}{d\tau^2}\chi(\tau)+m\omega^2\,\chi(\tau)-\lambda^2\int_{0}^{\infty}\!d\tau'\;G_{\textsc{r}}^{(\phi)}(z(\tau),z(\tau'))\,\chi(\tau')=\lambda\,\phi_{\text{h}}(z(\tau))\,.
\end{equation}
Next, we will evaluate the integral expression covariantly along the worldline of the detector to obtain a covariant formulation of the internal dynamics.

The retarded field $\phi_{\text{ret}}(x)$ generated by the internal dynamics at an arbitrary spacetime point $x$ is given by
\begin{align}
	\phi_{\text{ret}}(x)&=\lambda\int_{0}^{\infty}\!d\tau'\;G_{\textsc{r}}^{(\phi)}(x,z(\tau'))\,\chi(\tau')=\frac{1}{4\pi}\frac{\mathfrak{d}(\tau)}{r(\tau)}\biggr|_{\text{ret}}\,,\label{E:gbwore}
\end{align}
where $\mathfrak{d}(\tau)=\lambda\,\chi(\tau)$, and $R^{\mu}(\tau)=x^{\mu}-z^{\mu}(\tau)$ is the separation vector. The subscript ``ret" indicates that the quantity is evaluated at the retarded proper time $\tau{_{\text{ret}}}$, which is defined as the proper time in the past when the lightcone from the worldline intersects the observation point $x$. Mathematically, $\tau_{\text{ret}}$ is the unique root of the past light-cone condition $R_{\mu}(\tau_{\text{ret}})R^{\mu}(\tau_{\text{ret}})=0$. The spacelike invariant retarded distance $r$ is defined by $r(\tau_{\text{ret}})=R_{\mu}(\tau_{\text{ret}})u^{\mu}(\tau_{\text{ret}})$, where $u^{\mu}$ is the four-velocity along the worldline.

From the results in Appendix~\ref{S:ejheohti}, we arrive at the regularized retarded field on the worldline,
\begin{equation}
	\phi_{\text{ret}}(z(\tau))=\frac{\lambda}{4\pi}\frac{\chi(\tau)}{\epsilon}-\frac{\lambda}{4\pi}\,\frac{d}{d\tau}\chi(\tau)\,.
\end{equation}
Substituting this regularized field back into the effective equation of motion, Eq.~\eqref{E:eituef}, for the internal degree of freedom, we obtain
\begin{equation}\label{E:etdgbd}
	\ddot{\chi}(\tau)+\frac{\lambda^{2}}{4\pi m}\,\dot{\chi}(\tau)+\biggl[\omega^2-\frac{\lambda^{2}}{4\pi m\,\epsilon}\biggr]\,\chi(\tau)=\frac{\lambda}{m}\,\phi_{\text{h}}(z(\tau))\,.
\end{equation}
Note that the divergent shift $\delta\omega^{2}=-\dfrac{\lambda^{2}}{4\pi m\,\epsilon}$ is a scalar that is entirely independent of the detector's mechanical motion or acceleration history. If we had not employed a covariant regularization scheme, for example, by using a non-covariant spatial cutoff or prematurely expanding in the non-relativistic limit, this divergent shift would undesirably be included in the mechanical kinematics, picking up spurious, trajectory-dependent terms, such as contributions proportional to $\bm{v}\cdot\bm{a}$, as in Eq.~\eqref{E:dbskgs}. Since the fully covariant point-splitting method ensures that the divergence depends exclusively on the system's structural parameters $\lambda$, $m$, $\epsilon$, and is completely decoupled from the dynamical state of the detector, it acts as a constant shift to the bare oscillator rather than as a kinematic force. This allows us to absorb it by defining a finite, time-independent renormalized oscillating frequency 
\begin{equation}
	\omega^{2}_{\text{ren}}=\omega^{2}-\frac{\lambda^{2}}{4\pi m\,\epsilon}\,.
\end{equation}
In terms of the proper time $\tau$, the effective equation of motion for the internal degree of freedom reduces to the exact form of a standard, constant-coefficient driven harmonic oscillator,
\begin{equation}\label{E:foeue}
    \ddot{\chi}(\tau)+2\bar{\gamma}\,\dot{\chi}(\tau)+\omega_{\text{ren}}^2\chi(\tau)=\frac{\lambda}{m}\,\phi_{\text{h}}(z(\tau))\,,
\end{equation}
where we have identified the radiation damping constant $\bar{\gamma}$ from Eq.~\eqref{E:etdgbd} as
\begin{equation}
    \bar{\gamma}=\frac{\lambda^{2}}{8\pi m}\,.
\end{equation}

Let us compute the power of the damping term in this equation of motion. The quantity 
\begin{equation}
	\mathcal{P}_{\gamma}=-2m\bar{\gamma}\,\biggl(\frac{d\chi}{d\tau}\biggr)^{2}\,
\end{equation}
defines the proper power of damping, which is a Lorentz scalar. Physically, it represents the rate of energy transfer or the work done by the damping force as measured in the instantaneous rest frame of the detector, parameterized by its own proper time $\tau$. To connect this with the power measured in the laboratory frame, we first note that energy transforms as the time component of a four-vector. Since the radiation is emitted isotropically in the rest frame, the proper momentum change $d\mathcal{\bm{p}}$ is zero. Thus, the transformation of energy reduces to
\begin{equation*}
    dE^{(\text{lab})}=\gamma_{\bm{v}}\,\bigl(d\mathcal{E}+\bm{v} \cdot d\mathcal{\bm{p}}\bigr)\to\gamma_{\bm{v}}\,d\mathcal{E}_{\gamma}\,.
\end{equation*}
That is, the energy shift in the laboratory frame is simply scaled by the Lorentz factor $\gamma_{\bm{v}}(t)$. Consequently, the dissipative power in the laboratory frame is
\begin{equation}
	P_{\gamma}^{(\text{lab})}=\frac{dE_{\gamma}^{(\text{lab})}}{dt}=\gamma_{\bm{v}}\,\frac{d\mathcal{E}_{\gamma}}{dt}=\frac{d\mathcal{E}_{\gamma}}{d\tau}=\mathcal{P}_{\gamma}\,.
\end{equation}
Applying the time dilation relation $d\tau=\gamma_{\bm{v}}^{-1} dt$, we can rewrite this invariant power in terms of the coordinate time derivative, which we denote as $\dot{\chi}=d\chi/dt$,
\begin{equation}
	P_{\gamma}^{(\text{lab})}=-2m\bar{\gamma}\,\biggl(\frac{d\chi}{d\tau}\biggr)^{2}=-2m\bar{\gamma}\,\gamma_{\bm{v}}^{2}\,\dot{\chi}^{2}\,.
\end{equation}
This reveals that the dissipative radiated power is amplified by the mechanical motion of the detector via the $\gamma_{\bm{v}}^{2}$ factor.

The general solution to the effective equation of motion for the internal degree of freedom is given by
\begin{equation}
	\chi(\tau)=\chi_{\text{h}}(\tau)+\lambda\int_{0}^{\infty}\!d\tau'\;G_{\textsc{r}}^{(\chi)}(\tau,\tau')\,\phi_{\text{h}}(z(\tau'))\,,
\end{equation}
where the retarded Green's function $G_{\textsc{r}}^{(\chi)}$ takes the proper-time translation-invariant form of a standard driven damped harmonic oscillator,
\begin{equation}
	G_{\textsc{r}}^{(\chi)}(\tau,\tau')=\theta(\tau-\tau')\,\frac{1}{m\Omega}\,e^{-\bar{\gamma}(\tau-\tau')}\,\sin\Omega(\tau-\tau')\,,
\end{equation}
with the damped frequency defined as $\Omega=\sqrt{\omega_{\text{ren}}^{2}-\bar{\gamma}^{2}}$.  This integral representation explicitly shows that the driven portion of the internal dynamics is a functional of the ambient free field $\phi_{\text{h}}(z(\tau))$ evaluated along the worldline of the detector. The homogeneous solution $\chi_{\text{h}}(\tau)$ describes the transient, free evolution from the initial internal state, defined by $\chi_{0}=\chi(0)$ and $\dot{\chi}_{0}=d\chi(\tau)/d\tau\,\Big|_{\tau=0}$,
\begin{equation}
	\chi_{\text{h}}(\tau)=d_{1}(\tau)\,\chi_{0}+d_{2}(\tau)\,\dot{\chi}_{0}\,,
\end{equation}
where the fundamental solutions are
\begin{align}
	d_{1}(\tau)&=e^{-\bar{\gamma}\tau}\,\Bigl[\cos\Omega\tau+\frac{\bar{\gamma}}{\Omega}\,\sin\Omega\tau\Bigr]\,,&d_{2}(\tau)&=\frac{1}{\Omega}\,e^{-\bar{\gamma}\tau}\,\sin\Omega\tau\,.
\end{align}
Since these fundamental solutions decay exponentially, the initial conditions are eventually forgotten. Thus, the late-time internal dynamics are completely determined by the mechanical motion navigating through the free field. Note that we have synchronized the clocks such that $t=0$ and $\tau=0$ coincide at the exact spacetime event where the interaction is switched on. For any subsequent time $t>0$, the laboratory coordinate time $t$ at a given proper time $\tau$ is dictated by the detector's history. This is obtained by integrating the Lorentz factor along the worldline,
\begin{equation}
	t(\tau)=\int_{0}^{\tau}\!d\tau'\;\gamma_{\bm{v}}(\tau')\,.
\end{equation}

\section{Relativistic covariant dynamics of the mechanical dof}\label{S:otgyor}

Finally, we turn to the classical dynamics of the mechanical or external degree of freedom. The mechanical degree of freedom does not enter the idf oscillator  Lagrangian directly. However, the idf oscillator still affects the detector's motion because the proper time used to describe its dynamics depends on the detector's trajectory. This added dependence on the trajectory (on top of  the field being evaluated at the detector's position) in this interacting tripartite system is one easily forgotten factor which contributes to the flaws of a nonrelativistic description.

We start with the full relativistic action of the detector, comprising both its internal and mechanical degrees of freedom
\begin{align}\label{E:dnbete}
    S_{\text{detector}}=\int\!d\tau\;\biggl\{-\biggl[M+V(z)\biggr]+\biggl[\frac{m}{2}\,\chi'^2-\frac{m\omega^2}{2}\,\chi^2\biggr]+\lambda\,\chi\,\phi(z)\biggr\}
\end{align}
We wish to vary this action with respect to $z^{\mu}$ to get the equation of motion. However, because the proper time $\tau$ is not an independent variable and depends on the trajectory $z^{\mu}$ itself, we must re-parameterize the worldline using an arbitrary worldline parameter $\varsigma$. We denote the coordinate four-velocity with respect to $\varsigma$ by a prime, $z'^{\mu}=\dfrac{dz^{\mu}}{d\varsigma}$. At the end of the calculation, we will identify the parameter $\varsigma$ back with the physical proper time $\tau$. The proper time differential is given by $d\tau=\sqrt{z'^2}\,d\varsigma$, and the proper time derivative of the internal degree of freedom transforms to
\begin{equation}
    \frac{d\chi}{d\tau}=\frac{\chi'}{\sqrt{z'^2}}\,.
\end{equation}
This allows us to rewrite the action, Eq.~\eqref{E:dnbete}, entirely in terms of $\varsigma$,
\begin{align}
    S_{\text{detector}}=\int\!d\varsigma\;\sqrt{z'^2}\,\biggl\{-\biggl[M+V(z)\biggr]+\biggl[\frac{m}{2}\,\biggl(\frac{\chi'}{\sqrt{z'^2}}\biggr)^2-\frac{m\omega^2}{2}\,\chi^2\biggr]+\lambda\,\chi\,\phi(z)\biggr\}\,.
\end{align}
We then perform the variations of this action to identify the covariant equation of motion for $z^{\mu}$.

Adding Eqs.~\eqref{E:ngsr}, \eqref{E:ndfjet}, and \eqref{E:jetujet} together, we find that the variation of the detector's total action is given by
\begin{align}
    \delta S_{\text{detector}}&=\int\!d\tau\;\biggl[M\,\ddot{z}_{\mu}+\frac{d}{d\tau}\biggl\{\biggl[V(z)+\biggl(\frac{m}{2}\,\dot{\chi}^2+\frac{m\omega^2}{2}\,\chi^2\biggr)-\lambda\,\chi\,\phi(z)\biggr]\,\dot{z}_{\mu}\biggr\}\biggr.\notag\\
    &\qquad\qquad\qquad\qquad\qquad\qquad\qquad\qquad\qquad\qquad\qquad-\biggl.\partial_{\mu}V(z)+\lambda\,\chi\,\partial_{\mu}\phi(z)\biggr]\,\delta z^{\mu}\,.
\end{align}

\subsection{Proper-time dependent mass}

If we introduce the time-dependent quantity
\begin{equation}
    M(\tau)=M+V(z)+\biggl(\frac{m}{2}\,\dot{\chi}^2+\frac{m\omega^2}{2}\,\chi^2\biggr)-\lambda\,\chi\,\phi(z)\,,
\end{equation}
then the principle of stationary action, $\delta S_{\text{detector}}=0$, yields the equation of motion for the mechanical degree of freedom,
\begin{equation}\label{E:bifer}
    \frac{d}{d\tau}\Bigl[M(\tau)\,\dot{z}_{\mu}\Bigr]=\partial_{\mu}V(z)-\lambda\,\chi\,\partial_{\mu}\phi(z)\,.
\end{equation}
The expression $P^{\mu}=M(\tau)\,\dot{z}^{\mu}$ functions as the generalized four-momentum, with $M(\tau)$ acting as the dynamic inertia of the detector. Since the detector exchanges energy with both the internal oscillator and the scalar field, its inertia is no longer restricted to the bare rest mass $M$. These internal energies contribute dynamically to the mass, a direct manifestation of mass-energy equivalence. It is also worth mentioning that the additional contributions to the mass from Eqs.~\eqref{E:ngsr}, \eqref{E:ndfjet}, and \eqref{E:jetujet}, as a consequence of the $z^{\mu}$ dependence of the proper time, are absent in a treatment initiated from the nonrelativistic action. These terms not only complete the mass-energy relation in relativistic kinematics, they also have a profound impact on dynamics.

Taking the derivative of $M(\tau)$ with respect to $\tau$ and factoring out $\dot{\chi}$, we get
\begin{align}
    \frac{dM(\tau)}{d\tau}=\dot{z}^{\mu}\partial_{\mu}V+\bigl(m\,\ddot{\chi}+m\omega^2\,\chi-\lambda\,\phi\bigr)\,\dot{\chi}-\lambda\,\chi\,\dot{z}^{\mu}\partial_{\mu}\phi=\dot{z}^{\mu}\partial_{\mu}V-\lambda\,\chi\,\dot{z}^{\mu}\partial_{\mu}\phi\,,
\end{align}
where we have used the internal equation of motion, Eq.~\eqref{E:wygrwsf}, to set the term in parentheses to zero. We can then expand the equation of motion, Eq.~\eqref{E:bifer}, and find the acceleration force
\begin{equation}\label{E:fbirtr}
    M(\tau)\,a_{\mu}=-\dot{z}_{\mu}\dot{z}_{\nu}\partial^{\nu}V+\lambda\,\chi\,\dot{z}_{\mu}\dot{z}_{\nu}\,\partial^{\nu}\phi+\partial_{\mu}V(z)-\lambda\,\chi\,\partial_{\mu}\phi(z)=P_{\mu\nu}\bigl(\partial^{\nu}V-\lambda\,\chi\,\partial^{\nu}\phi\bigr)
\end{equation}
where $P_{\mu\nu}=\eta_{\mu\nu}-\dot{z}_{\mu}\dot{z}_{\nu}$ is the transverse projection tensor orthogonal to the four-velocity. This projection mathematically guarantees that only the components of the forces perpendicular to the four-velocity can change the four-acceleration, preserving the normalization of the four-velocity. In other words, if we define the four-force as the derivative of the four-momentum, then in general, the force can point in any spacetime direction, not necessarily spacelike. In order to identify the four-acceleration, we need to project the four-force onto the direction normal to $u^{\mu}$ by applying the projection operator $P^{\mu\nu}$,
\begin{align}\label{E:goeiuo}
    \mathcal{F}^{\mu}&=\frac{d\mathcal{P}^{\mu}}{d\tau}=Ma^{\mu}+\dot{M}u^{\mu}\,,&&\Rightarrow&a^{\mu}&=\frac{1}{M}\,P^{\mu}{}_{\nu}\,\mathcal{F}^{\nu}\,,
\end{align}
where $\mathcal{P}^{\mu}=Mu^{\mu}$.

\subsection{Divergences and Renormalization}

The equation of motion for the mechanical degree of freedom is a functional of the full field $\phi$, which is the sum of the homogeneous free field $\phi_{\text{h}}$ and the retarded self-field $\phi_{\text{ret}}$ generated by the internal dynamics. Because $\phi_{\text{ret}}$ formally diverges on the detector's own worldline, we must regularize this divergent component so it can be absorbed into the renormalized parameters. To maintain a covariant framework, we employ a covariant cutoff regularization method.

With details given in Appendix~\ref{S:eothgbd} and from Eq.~\eqref{E:dfdf}, we finally arrive at the total self-field gradient
\begin{align}\label{E:btidfd}
	\partial^{\mu}\phi_{\text{reg}}(x)&=-\frac{1}{16\pi}\,\mathfrak{d}\,u^{\mu}\,\Lambda^{2}+\biggl[\frac{1}{8\pi}\,\mathfrak{d}\,a^{\mu}+\frac{1}{4\pi}\,\dot{\mathfrak{d}}\,u^{\mu}\biggr]\,\Lambda\notag\\
    &\qquad\qquad\qquad+\biggl[-\frac{1}{12\pi}\,\mathfrak{d}\,a^{2}\,u^{\mu}-\frac{1}{12\pi}\,\mathfrak{d}\,b^{\mu}-\frac{1}{4\pi}\,\dot{\mathfrak{d}}\,a^{\mu}-\frac{1}{4\pi}\,\ddot{\mathfrak{d}}\,u^{\mu}\biggr]+\mathcal{O}(\Lambda^{-1})\,,
\end{align}
{where $b^{\mu}$ is the shorthand notation for $\dot{a}^{\mu}=\dddot{z}^{\mu}$.} We now evaluate the transverse self-force exerted on the detector by its own regularized field, $-\mathfrak{d}\, P_{\mu\nu}\partial^{\nu}\phi_{\text{reg}}$, to extract the four-acceleration according to Eq.~\eqref{E:goeiuo}. Applying the projection operator $P_{\mu\nu}=\eta_{\mu\nu}-u_{\mu} u_{\nu}$ to the regularized field gradient $\partial^{\nu}\phi_{\text{reg}}$, Eq.~\eqref{E:btidfd}, completely eliminates the most severe $\mathcal{O}(\Lambda^2)$ divergence, so the surviving terms from the projection are
\begin{equation}
    P_{\mu\nu}\partial^{\nu}\phi_{\text{reg}}=\frac{\Lambda}{8\pi}\,\mathfrak{d}\,a_{\mu}-\frac{1}{4\pi}\,\dot{\mathfrak{d}}\,a_{\mu}-\frac{1}{12\pi}\,\mathfrak{d}\,\bigl(b_{\mu}+a^2u_{\mu}\bigr)+\mathcal{O}(\Lambda^{-1})\,.
\end{equation}
Multiplying by $-\mathfrak{d}$ and retaining terms up to $\mathcal{O}(\Lambda^0)$, the projected self-force takes the asymptotic form
\begin{equation}
    -\mathfrak{d}\,P_{\mu\nu}\partial^{\nu}\phi_{\text{reg}}=-\frac{\Lambda}{8\pi}\,\mathfrak{d}^2\,a_{\mu}+\frac{1}{4\pi}\,\mathfrak{d}\dot{\mathfrak{d}}\,a_{\mu}+\frac{1}{12\pi}\,\mathfrak{d}^2\,\bigl(b_{\mu}+a^2u_{\mu}\bigr)+\mathcal{O}(\Lambda^{-1})\,,
\end{equation}
where $\mathfrak{d}(\tau)=\lambda\,\chi(\tau)$ accounts for internal dynamics.

{We note that the projection entirely annihilates the original terms proportional to $u^{\mu}$ in Eq.~\eqref{E:btidfd}. The newly emergent, nonlinear kinematic term $a^2u_{\mu}$ arises naturally from the transverse projection of $b_{\mu}$, because the normalization condition of the four velocity implies
\begin{align}
    u_{\mu}u^{\mu}&=1\,,&&\implies &a_{\mu}u^{\mu}&=0\,,&&\implies &a^2+b_{\mu}u^{\mu}&=0\,.
\end{align}
}

{The resulting terms on the right-hand side logically separate into inertial corrections and dissipative forces. The first term is a linear $\mathcal{O}(\Lambda)$ divergence aligned with the acceleration. It is conventionally identified as the divergent mass shift $\delta M$. The second term acts as an additional, finite mass correction emerging purely from the time-dependence of the internal degree of freedom. Within this classical framework, it is mathematically straightforward to move both of these terms to the left-hand side of Eq.~\eqref{E:fbirtr} and formally absorb them into a time-dependent, renormalized effective mass \footnote{Including dynamic variables into the definition of a generalized or effective mass seems like an unavoidable outcome of a relativistic covariant treatment.  This measure is not entirely unfamiliar, renormalization in quantum field theory is a routinely implemented. The old concept of inertia phrased in Newton's First Law may need some augmentation. Afterall, neither relativity or quantum has not quite arrived then.}.  Finally, the remaining non-conservative $\mathcal{O}(\Lambda^0)$ term is precisely the scalar-field counterpart to the classical Abraham-Lorentz-Dirac (ALD) radiation reaction.}

The equation of motion Eq.~\eqref{E:fbirtr} is now given by
\begin{equation}\label{E:ebdigdg}
    M_{\text{ren}}(\tau)\,a_{\mu}=P_{\mu\nu}\bigl(\partial^{\nu}V-\lambda\,\chi\,\partial^{\nu}\phi_{\text{h}}\bigr)+\frac{\lambda^2}{12\pi}\,\chi^2\,\bigl(b_{\mu}+a^2u_{\mu}\bigr)\,,
\end{equation}
where
\begin{equation}\label{E:dbierue}
    M_{\text{ren}}(\tau)=M+\frac{\lambda^2\Lambda}{8\pi}\,\chi^2+V(z)+\biggl(\frac{m}{2}\,\dot{\chi}^2+\frac{m\omega^2}{2}\,\chi^2\biggr)-\lambda\,\chi\,\phi(z)-\frac{\lambda^2}{4\pi}\,\chi\dot{\chi}\,,
\end{equation}
denotes the renormalized mass. In this form, Eq.~\eqref{E:ebdigdg} appears almost identical to the equation of motion of a structureless point charge, except that in this case the renormalized mass $M_{\text{ren}}(\tau)$ is still time dependent. There are a few subtle but interesting points. The divergent mass shift $\delta M=\dfrac{\lambda^2\Lambda}{8\pi}\,\chi^2$ is actually frequency renormalization in disguise. Lumping this divergent mass shift into the internal-energy contribution of $\chi$ to $M_{\text{ren}}(\tau)$, we have
\begin{equation}
    \frac{m}{2}\,\dot{\chi}^2+\frac{m\omega^2}{2}\,\chi^2+\frac{\lambda^2\Lambda}{8\pi}\,\chi^2=\frac{m}{2}\,\dot{\chi}^2+\frac{m}{2}\,\biggl(\omega^2+\frac{\lambda^2\Lambda}{4\pi m}\biggr)\,\chi^2=\frac{m}{2}\,\dot{\chi}^2+\frac{m\omega^2_{\text{ren}}}{2}\,\chi^2\,.
\end{equation}
However, the sign of $\delta\omega^2$ does not match the result in Eq.~\eqref{E:etdgbd}. This disparity results from neglecting the contribution from the term $-\lambda\,\chi\,\phi(z)$ in $M_{\text{ren}}(\tau)$, where the full field $\phi$ also contains a divergent contribution.

Following the same regularization scheme, the regularized field $\phi_{\text{reg}}$ at the location of the detector is given by
\begin{align}
	\phi_{\text{reg}}(z(\tau))=\frac{\Lambda}{4\pi}\int_{-\infty}^{\tau}\!d\tau'\;\mathfrak{d}(\tau')\,\theta(\sigma'^{2})\,\frac{J_{1}(\Lambda\,\sigma')}{\sigma'}&=\frac{\Lambda}{4\pi}\int_{0}^{\infty}\!ds\;\mathfrak{d}(\tau-s)\,\frac{J_{1}(\Lambda\,\sigma(s))}{\sigma(s)}\notag\\
	&=\frac{1}{4\pi}\int_{0}^{\infty}\!dy\;\biggl\{\mathfrak{d}\,\frac{J_{1}(y)}{y}\,\Lambda-\dot{\mathfrak{d}}\,J_{1}(y)+\mathcal{O}(\Lambda^{-1})\biggr\}\notag\\
	&=\frac{\mathfrak{d}}{4\pi}\,\Lambda-\frac{\dot{\mathfrak{d}}}{4\pi}\,,
\end{align}
where $\sigma^{2}>0$ because the detector follows a timelike trajectory. Thus, we have the full field expanded as
\begin{equation}\label{E:oebdere}
    \phi=\phi_{\text{h}}+\frac{\mathfrak{d}}{4\pi}\,\Lambda-\frac{\dot{\mathfrak{d}}}{4\pi}\,.
\end{equation}
Substituting this into $M_{\text{ren}}(\tau)$ gives
\begin{align}\label{E:ldbnoeur}
     M_{\text{ren}}(\tau)&=M-\frac{\lambda^2\Lambda}{8\pi}\,\chi^2+V(z)+\biggl(\frac{m}{2}\,\dot{\chi}^2+\frac{m\omega^2}{2}\,\chi^2\biggr)-\lambda\,\chi\,\phi_{\text{h}}(z)\notag\\
     &=M+V(z)+\biggl(\frac{m}{2}\,\dot{\chi}^2+\frac{m\omega_{\text{ren}}^2}{2}\,\chi^2\biggr)-\lambda\,\chi\,\phi_{\text{h}}(z)\,,
\end{align}
where the squared renormalized frequency is now consistently given by
\begin{equation}
    \omega_{\text{ren}}^2=\omega^2-\frac{\lambda^2\Lambda}{4\pi m}\,.
\end{equation}
{Then the time-dependent divergence $\delta M$ in the acceleration entirely vanishes, having been absorbed into the finite physical oscillating frequency of the internal degree of freedom. A more implicit consequence is that the last term $-\dfrac{\lambda^2}{4\pi}\,\chi\dot{\chi}$ in $M_{\text{ren}}$, Eq.~\eqref{E:dbierue}, is also completely canceled by the finite contribution $-\dfrac{\dot{\mathfrak{d}}}{4\pi}$ in $\phi$ or $\phi_{\text{reg}}$.}

The final form of the equation of motion is 
\begin{equation}\label{E:dnldger}
    M_{\text{ren}}(\tau)\,a_{\mu}=P_{\mu\nu}\bigl(\partial^{\nu}V-\lambda\,\chi\,\partial^{\nu}\phi_{\text{h}}\bigr)+\frac{\lambda^2}{12\pi}\,\chi^2\,\bigl(b_{\mu}+a^2u_{\mu}\bigr)\,,
\end{equation}
where $a^{\mu}=d^2z^{\mu}/d\tau^2$ and $b^{\mu}=d^3z^{\mu}/d\tau^3$. After frequency renormalization, the cutoff-dependent local terms are formally absorbed into the physical parameters, leaving only the finite scalar radiation-reaction force to act on the detector. The spurious artifacts that appeared earlier in the raw expression for the self-force are systematically eliminated and covariance preserved. Terms proportional to the four-velocity $u^{\mu}$ are completely annihilated by the transverse projection $P_{\mu\nu}$, while terms proportional to the four-acceleration $a^{\mu}$ are subsumed into the renormalized effective mass $M_{\text{ren}}(\tau)$, facilitated by mass-energy equivalence. These renormalization procedures, entirely absent in conventional nonrelativistic formulations, successfully secure classical covariance. Algebraically they do so by shifting the scalar field interaction into the definition of inertia itself.

{This mathematically elegant treatment based on relativity principles for a composite systems in a classical field  will be subjected to close scrutiny at another level when we deal with quantum fields. There, excitation of quantum fluctuations in the background field should be included in the interaction with the idf, and their resultant back-reaction on the dynamics of the  mechanical degrees of freedom  may reveal a different scenario.}

\subsection{The nonrelativistic limit and other issues}

Formally, the resulting equation of motion for the mechanical degree of freedom, Eq.~\eqref{E:dnldger}, closely resembles that of a structureless point scalar charge. This occurs because the field couples to the composite system exclusively through the effective source $\mathfrak{d}(\tau)=\lambda\,\chi(\tau)$ localized on the worldline. The field does not resolve the microscopic internal structure of the detector. Consequently, the internal dynamics manifest purely through the time-dependence of the effective charge and the renormalized inertia, while the local self-force retains the universal, covariant geometry associated with a pointlike monopole.

{To explicitly demonstrate this correspondence, we take the nonrelativistic limit of the radiation reaction term $b^{\mu}+a^2u^{\mu}$. Expanding the kinematic variables to their lowest non-vanishing orders in $\lvert\bm{v}\rvert\ll1$, we have
\begin{align}
    u^{\mu}&\approx(1+\frac{v^2}{2},\,\bm{v})\,,&a^{\mu}&\approx(\bm{v}\cdot\bm{a},\,\bm{a})\,,&a^2&\approx-\bm{a}^2\,,&b^{\mu}&\approx(\bm{a}^2+\bm{v}\cdot\bm{b},\,\bm{b})\,,
\end{align}
with $\bm{v}=d\bm{z}/dt$, $\bm{a}=d\bm{v}/dt$ and $\bm{b}=d\bm{a}/dt$. Substituting these into the radiation reaction term yields
\begin{equation}
    b^{\mu}+a^2u^{\mu}=(\bm{v}\cdot\bm{b},\,\bm{b}-\bm{a}^2\bm{v})\,.
\end{equation}
It is crucial to emphasize that the presence of the nonlinear spatial term $-\bm{a}^2\,\bm{v}$ does not signal  {``vacuum viscosity''} induced by the internal structure of the detector. Rather, it is a universal kinematic consequence of relativistic momentum conservation for any radiating detector, appearing identically in the Abraham-Lorentz-Dirac force for a standard electromagnetic point charge, and is thus not a unique feature of the scalar field. Since this term scales with the square of the detector's acceleration, it identically vanishes when the detector moves with uniform speed. Thus, within the classical regime, the internal degree of freedom merely modulates the amplitude of the standard radiation reaction without independently generating any drag during uniform motion. Clarifying this deterministic behavior is helpful for a theoretical transition to quantum field theory, where the interaction of the internal state with vacuum fluctuations will modify the dynamics of uniform motion, introducing effects that are entirely absent in the classical treatment. Since the vacuum expectation value vanishes, this induced drag cannot manifest at the mean-field level. It arises inherently as a fluctuation-driven phenomenon due to the dynamical correlations established among the internal subsystem, the detector's trajectory, and the quantum vacuum.}

\section{Summary and Discussions}\label{S:obyrw}

\subsection{Key findings}
In the preceding sections, in a fully relativistic, covariant formulation, we demonstrated how the internal microdynamics and the macroscopic mechanical motion of a composite, radiating system are intricately linked through a background free scalar field $\phi_{\text{h}}$. We also pointed out the inadequacies of an ab initio nonrelativistic treatment. The salient mathematical features and physical implications of this fully covariant treatment can be distilled into the following key findings:
\begin{enumerate}
    \item Treating the detector's center-of-mass and internal dynamics nonrelativistically while coupling them to an inherently relativistic field explicitly breaks covariance. This mismatch manifests as spurious, trajectory-dependent terms, such as the velocity-dependent damping and the non-inertial frequency modulation $\bm{v}\cdot\bm{a}$ seen in Eq.~\eqref{E:dbskgs}. If taken at face value, these artifacts incorrectly suggest that an unaccelerated detector experiences spontaneous deceleration (a classical ``vacuum viscosity") or parametric mechanical instabilities.
    \item When formulated covariantly using the proper time $\tau$, the rest mass of the detector is no longer a static parameter. The principle of stationary action reveals a dynamic, time-dependent effective mass $M_{\text{ren}}(\tau)$, as given in Eq.~\eqref{E:ldbnoeur}. This renormalized inertia properly absorbs the kinetic and potential energies of the internal degree of freedom, alongside the interaction energy with the free scalar field, a clear observation of the mass-energy equivalence at the macroscopic level.
    \item The fully covariant derivation proves that, at the classical level, there is no vacuum viscosity. An unaccelerated detector moving through a classical vacuum experiences no spontaneous velocity-dependent drag. The apparent deceleration derived in the nonrelativistic approximation is  exposed as a coordinate artifact rather than a genuine physical effect.
    \item At a more profound theoretical level, our derivation demonstrates how the system's microdynamics act as a guardrail for Lorentz invariance. Even though the internal oscillator continuously radiates and actively modifies the local field configuration (generating the retarded self-field $\phi_{\text{ret}}$), this dynamic back-reaction is prohibited from breaking spacetime symmetries. The covariant framework guarantees that the back-reaction of this modified field perfectly partitions into Lorentz-preserving structures exemplified in the following ways:
        \begin{itemize}[label=$\circ$]
            \item Spurious longitudinal forces, which would otherwise induce vacuum drag, are mathematically annihilated by the kinematic constraints of relativity via the transverse projection tensor $P_{\mu\nu}$.
            \item The $\mathcal{O}(\Lambda)$ divergent self-energy of the interaction is systematically absorbed and repackaged into the physical definitions of macroscopic inertia, $M_{\text{ren}}$, and internal frequency, $\omega_{\text{ren}}$.
            \item The surviving non-conservative term resolves precisely into the scalar analog of the Abraham-Lorentz-Dirac (ALD) radiation reaction force, $\dfrac{\lambda^2}{12\pi}\chi^2(b_{\mu}+a^2\,u_{\mu})$.
        \end{itemize}
\end{enumerate}

\subsection{A broad issue and some technical finesse with quantum fields} 

{Since dissipation by friction or viscosity is the central physical issue of interest here we first make a remark on how to identify dissipative terms by their parity under time reversal.  After that}, while the treatment of this composite system in a classical field is fresh on our minds, we mention here some subtle differences when we carry out a similar investigation based on quantum field theory where vacuum fluctuations need be included. 

{\paragraph{Time-reversal parity and the power signature of dissipation}
On a broader theoretical note, our derivation highlights a universal diagnostic link between time-reversal parity, mechanical power, and dissipation. Terms with even time-reversal parity (e.g., $\ddot{z}^{\mu}$ or $\dot{\chi}^2$) are conservative, typically manifesting as parameter renormalizations or reversible potentials. Odd-parity terms explicitly break time symmetry, yet they only represent genuine dissipation if they yield negative mechanical power (e.g., $P=\bm{F}\cdot\bm{v}<0$), as demonstrated by the internal damping ($2\bar{\gamma}\dot{\chi}$) and mechanical loss ($-a^2\bm{v}$). Recognizing this distinction clarifies the fundamental nature of radiation backreaction. As formally captured by the symmetric and antisymmetric field decomposition in Eq.~\eqref{E:vndof}, the self-force naturally segregates into these two distinct dynamical classes: conservative terms (originating from the bond field) that dynamically reshape the system's inertia, and dissipative terms (originating from the radiation field) that irreversibly carry away energy. Tracking both parity and power thus reveals the precise mechanism by which the classical system balances its energy budget, establishing a rigorous baseline before quantum vacuum fluctuations are introduced.}

\paragraph{What enters in the inertia? -- further renormalizations} 
{Examining the explicit form of the inertia in Eq.~\eqref{E:ldbnoeur}, we observe that $M_{\text{ren}}(\tau)$ comprises the bare rest mass, the external potential energy, the mechanical energy of the internal degree of freedom, and the scalar interaction energy. Two features are particularly noteworthy. First, it is the homogeneous free field, $\phi_{\text{h}}$, rather than the full field, that enters the inertia. Second, within this specific grouping, there appears to be no residual mass renormalization, which might naively imply that the bare rest mass need not be infinite. However, we must treat this classical finiteness with caution. When adapted to the framework of quantum field theory, even though the parameters associated with the internal degree of freedom have been renormalized, the mechanical energy and the interaction term $-\lambda\,\chi\,\phi_{\text{h}}$ will inevitably harbor their own quantum divergences, due to zero-point fluctuations. Consequently, further regularization procedures must be implemented to extract physically meaningful, finite values.} An interesting physical effect would be: If an Unruh-DeWitt detector accelerates, it would perceive the quantum vacuum as a thermal bath as its internal state $\chi$ gets excited. This implies that acceleration thermality would dynamically increase the detector's inertia. We hope to report on our study of these novel effects related to acceleration radiation from a UdW detector with mass in a later communication.

\paragraph{Quantum fluctuations of the detector's inertia} The presence of $\phi_{\text{h}}$ inside $M_{\text{ren}}$ inevitably induces quantum fluctuations in the detector's inertia itself. Even if the bare rest mass is treated classically, large vacuum field fluctuations can easily invalidate the positivity of the total mass, driving the system into runaway tachyonic instabilities \footnote{A tempting, though ultimately unsatisfactory, resolution might be to reconstruct the equation of motion by moving the interaction term $-\lambda\,\chi\,\phi_{\text{h}}\,a_{\mu}$ back to the right-hand side, treating it as an external driving force. However, this algebraic rearrangement simply trades one pathology for another: 1) by scaling with the acceleration, the force remains kinematically indistinguishable from inertia, 2) it transforms the equation of motion into an iterative form, and 3) it introduces a highly pathological multiplicative noise.}.

\paragraph{Special nature of a scalar field} used in this model investigation versus the more realistic vector (electromagnetism) and tensor fields (gravity). It appears there is some tension between canonical quantum stability and classical scalar interactions. Standard vector and tensor fields only alter a detector's inertia if the detector possesses an extended internal structure, like a dipole, or if the field partakes in shaping the spacetime metric itself, such as in gravity. In contrast, the scalar field is unique because it naturally evades these structural requirements. Its most basic interaction acts as a direct, spacetime-dependent shift to the rest mass itself, inextricably linking the detector's inertia to the volatile fluctuations of the quantum vacuum.

\subsection{Summary Messages} 

We first mention two points regarding the inadequacies of the nonrelativistic treatment, and the necessity of a fully relativistic framework even for treating systems functioning in the nonrelativistic regime.  We then discuss two bigger issues: 1) the difference between our dynamical approach from the kinematical (symmetry) approach invoking Lorentz invariance. 2) the differences between results obtained based on a closed system versus an open system viewpoint. 

\paragraph{Nonrelativistic treatment is an approximation which may give wrong answers}

We have seen that nonrelativistic and  relativistic treatments give qualitative different descriptions: there is vacuum viscosity in the former, but not the latter. In practical terms, in atomic physics where nonrelativistic treatments of atom-field interaction is prevalent when the atom is not fast moving or when radiation is not of the main concern. Yet in theoretical terms a complete description requires a fully relativistic treatment where one sees no vacuum viscosity. The calculations which led to these conclusions are both valid, which makes this issue even more interesting. It is the framework which should be compared. We should find out the essential differences and identify the inadequacies in the conclusions drawn from a nonrelativistic calculation. Methodologically, the flaw originates from taking an approximation to be the whole truth.  Under this approximation appears the viscous force term, which is not there in a fully relativistic calculation.  Theoretically, one can identify the pivotal point being the mass-energy relation in special relativity. As we have seen, allowing for a time-dependent effective mass in the equation of motion, which we  refer to as ``relativistic inertia",  resolves the issue and safeguards the validity of Newton's first law.  

\paragraph{Relativistic is more than just fast motion, it is a whole new theory}  

Overall, what we have learned are the following:  a)  ``relativistic” is more than just ``fast-moving”,  there is a whole theory behind it, which entails absolute space and time versus spacetime, proper time versus coordinate time, Lorentz invariance, mass-energy equivalence etc. It is Newton against Minkowski and Lorentz. b) When a field is involved, special relativity need to be invoked.  Radiative processes are intrinsically relativistic.  c) Always perform a fully relativistic calculation, only after the complete results are obtained should one make a nonrelativistic approximation. If an \textit{ab initio} nonrelativistic calculation gives a different answer, {trust the relativistic results and take the nonrelativistic limit.}   

\paragraph{Symmetry arguments, kinematical vs dynamical workings}

In the Introduction we mention that a keen reader with a background in particle physics or field theory may, after merely reading the title of this paper, provide an answer, that the moving object should not experience any damping force, by invoking Lorentz invariance   with the observation that most objects are composites.  We appreciate this simple and elegant approach, but want to see the inner workings of the three parties involved in this problem, specifically, how the interaction between the object's internal degrees of freedom with an external field affects its motion.  Historically, plenty of work was devoted to how Lorentz invariance acts on the internal structures, or how the symmetries of the internal degrees of freedom fare with Lorentz invariance, e.g., \cite{Finkel,Nambu,Ji}. They are mostly based on group theory and algebraic methods. Here, we emphasize more on the dynamics, from the activities of the idf in its interaction with the field to the motion of the object. In a way it is closer in spirit to inquires often made in atomic-optical physics and quantum optomechanics.  That was also why we wanted to go into details in a nonrelativistic calculation and tried to  identify its shortcomings,  what could go wrong if one does not pay enough attention to the underlying principles from a `bigger' or `deeper' theory. The same warning of latent shortcomings in   current theories of quantum information which are based on quantum mechanics. Quantum information theories need be based on quantum field theory \cite{AHS}, which incorporates special relativity principles, to avert causality and covariance issues.

\vskip18pt
\noindent{\bf Acknowledgments} 
J.-T. Hsiang is supported by the National Science and Technology Council of Taiwan, R.O.C. under Grant No.~NSTC 113-2112-M-011-001-MY3.

\newpage

\appendix
\section{Evaluation of \texorpdfstring{$\phi_{\mathrm{ret}}(x)$}{phi\_ret(x)}}
\label{S:eouhegd}
Here we outline the derivation of $\phi_{\text{ret}}(x)$
\begin{equation*}
	\phi_{\text{ret}}(x) = \frac{\lambda}{2\pi}\int_{0}^{\infty}\!du\;\chi(t-u)\,\theta(u)\,\delta(\sigma-\varepsilon^{2})\,.
\end{equation*}
Essentially, we need to evaluate an integral of the form
\begin{equation*}
	\int_{0}^{\infty}\!du\;f(u)\,\delta(\sigma-\varepsilon^{2}) = \frac{f(u_{\varepsilon})}{\sigma'(u_{\varepsilon})}\,,
\end{equation*}
where $u_{\varepsilon}$ is the root of $\sigma(u_{\varepsilon})=\varepsilon^2$, given by
\begin{equation}\label{E:rotiro}
	u_{\varepsilon}=\frac{1}{\sqrt{A}}\,\varepsilon-\frac{B}{2A^{2}}\,\varepsilon^{2}+\biggl(\frac{5B^{2}}{8A^{\frac{7}{2}}} - \frac{C}{2A^{\frac{5}{2}}}\biggr)\,\varepsilon^{3}+\cdots\,.
\end{equation}
Here $f(u)=\chi(t-u)$. Since $\varepsilon\to0$ and consequently $u_{\varepsilon}\to0$, we Taylor expand $f(u)$ and $\sigma(u)$ respectively as
\begin{align*}
	f(u)&\simeq\chi(t)-u\,\dot{\chi}(t)+\frac{u^{2}}{2}\,\ddot{\chi}(t)-\frac{u^{3}}{6}\,\dddot{\chi}(t)+\dots\,,\\
	\sigma(u)&\simeq A\,u^{2}+B\,u^{3}+C\,u^{4}+\cdots\,,
\end{align*}
such that
\begin{align}
	\frac{f(u_{\varepsilon})}{\sigma'(u_{\varepsilon})}&=\frac{c_{0}}{2\sqrt{A}}\frac{1}{\varepsilon}+\frac{A\,c_{1}-B\,c_{0}}{2A^{2}} + \mathcal{O}(\varepsilon)\,,
\end{align}
with
\begin{align*}
	c_{0}&=\chi(t)\,,  &c_{1}&=-\dot{\chi}(t)\,,&A(t)&=1-\bm{v}^{2}(t)\,,  &B(t)&=\bm{v}(t)\cdot\bm{a}(t)\,.
\end{align*}
Thus, the divergent part of $\phi_{\text{ret}}$ is
\begin{align}
	\phi_{\text{ret}}^{(\text{fr})}&=\frac{\lambda}{2\pi\varepsilon}\biggl(\frac{c_{0}}{2\sqrt{A}}\biggr)=\frac{\lambda}{4\pi\varepsilon}\,\gamma_{\bm{v}}(t)\,\chi(t)\,,
\end{align}
where $\gamma_{\bm{v}}= (1-\bm{v}^2)^{-\frac{1}{2}}$ is the Lorentz factor, and we note the relation $\dot{\gamma}_{\bm{v}}=\gamma^{3}_{\bm{v}}\,(\bm{v}\cdot\bm{a})$. The finite part of $\phi_{\text{ret}}$ is given by
\begin{align}
	\phi_{\text{ret}}^{(\text{finite})}&=\frac{\lambda}{2\pi}\biggl(\frac{A\,c_{1}-B\,c_{0}}{2A^{2}}\biggr)=-\frac{\lambda}{4\pi}\,\gamma_{\bm{v}}^{2}(t)\,\dot{\chi}(t)-\frac{\lambda}{4\pi}\,\gamma^{4}_{\bm{v}}(t)\,\chi(t)\,\bigl[\bm{v}(t)\cdot\bm{a}(t)\bigr]\,.
\end{align}
Expanding to leading non-trivial order in the nonrelativistic limit ($v \ll 1$), the finite part becomes
\begin{align}
	\phi_{\text{ret}}^{(\text{finite})}&\simeq-\frac{\lambda}{4\pi}\,\bigl[1+\bm{v}^{2}(t)\bigr]\,\dot{\chi}(t)-\frac{\lambda}{4\pi}\,\chi(t)\,\bigl[\bm{v}(t)\cdot\bm{a}(t)\bigr]\,.
\end{align}

\section{Evaluation of the gradient of the retarded field}\label{S:eohodghfg}
We want to evaluate the gradient of the nonlocal expression in Eq.~\eqref{E:doghoer}, involving the retarded Green's function
\begin{equation*}
    \int_{0}^{t}\!dt'\;\bm{\nabla}_{\bm{z}}G_{\textsc{r}}^{(\phi)}(t,\bm{z}(t)\,;\,t',\bm{z}(t'))\,\chi(t')\,.
\end{equation*}
We shift away from the worldline to $\bm{x}$ for the moment by introducing a tiny observation/cutoff distance $\bm{\epsilon}$ (referring to Fig.~\ref{Fi:lcReg} for geometry),
\begin{equation}
	\lambda^2\chi(t)\int_0^t\!du\;\bm{\nabla}_{\bm{z}} G_{\textsc{r}}^{(\phi)}(t,\bm{x};t',\bm{z}(t'))\,\bigg|_{\bm{x}=\bm{z}(t)+\bm{\epsilon}}\,\chi(t-u)\,.
\end{equation}
where $t'=t-u$. The gradient of the retarded Green's function gives
\begin{align}
	\bm{\nabla}_{\bm{z}} G_{\textsc{r}}^{(\phi)}(t,\bm{x};t',\bm{z}(t'))\,\bigg|_{\bm{x}=\bm{z}(t)+\bm{\epsilon}}&=\frac{1}{2\pi}\,\theta(u)\,\bm{\nabla}_{\bm{z}}\delta(u^{2}-\lvert\bm{r}\rvert^{2})=-\frac{1}{\pi}\,\theta(u)\,\bm{r}(u)\,\delta'(u^{2}-\lvert\bm{r}\rvert^{2})\,,
\end{align}
where the spatial distance vector $\bm{r}(u)=\bm{x}-\bm{z}(t-u)=\bm{\epsilon}+\bigl[\bm{z}(t)-\bm{z}(t-u)\bigr]$ links $\bm{x}$ to the retarded point $\bm{z}(t')$. We Taylor-expand the detector's past trajectory, $\bm{z}(t-u)$, around the present time $t$
\begin{equation}
	\bm{r}(u)=\bm{\epsilon}+\biggl[u\,\dot{\bm{z}}(t)-\frac{u^2}{2}\,\ddot{\bm{z}}(t)+\frac{u^3}{6}\,\dddot{\bm{z}}\biggr]\,.
\end{equation}
Since $\bm{\epsilon}$ is a small spacelike point-splitting vector with $\lvert\bm{\epsilon}\rvert=\epsilon$, the squared spatial interval $\lvert\bm{r}\rvert^{2}$ inside the delta function can simply be written as $\lvert\bm{r}(u)\rvert^2=\epsilon^2+X(u)$, where
\begin{align*}
	X(u)&=2u\,\bm{\epsilon}\cdot\dot{\bm{z}}(t)+u^2\,\Bigl[\dot{\bm{z}}^{2}(t)-\bm{\epsilon}\cdot\ddot{\bm{z}}(t)\Bigr]+u^{3}\,\Bigl[\frac{1}{3}\,\bm{\epsilon}\cdot\dddot{\bm{z}}(t)-\dot{\bm{z}}(t)\cdot\ddot{\bm{z}}(t)\Bigr]+\mathcal{O}(u^4)\,.
\end{align*}
We retain only the first-order correction in $X(u)$. Thus, we can write the gradient of the retarded Green's function as
\begin{align}
	\bm{\nabla}_{\bm{z}} G_{\textsc{r}}^{(\phi)}(t,\bm{x};t',\bm{z}(t'))\,\bigg|_{\bm{x}=\bm{z}(t)+\bm{\epsilon}}&=-\frac{1}{\pi}\,\theta(u)\,\bm{r}(u)\,\delta'(u^{2}-\epsilon^2-X(u))\notag\\
	&=-\frac{1}{\pi}\,\theta(u)\,\bm{r}(u)\,\Bigl[\delta'(u^{2}-\epsilon^2)-X(u)\,\delta''(u^{2}-\epsilon^2)+\cdots\Bigr]\,,
\end{align}
and the integral expression in Eq.~\eqref{E:rmtrio} becomes
\begin{align}\label{E:ybgoe}
	\lambda^{2}\chi(t)\int_{0}^{t}\!dt'\;\bm{\nabla}_{\bm{z}}G_{\textsc{r}}^{(\phi)}(t,\bm{z}(t)\,;\,t',\bm{z}(t'))\,\chi(t')&=-\frac{\lambda^2}{\pi}\,\chi(t)\int_0^t\!du\;\theta(u)\,\bm{r}(u)\,\chi(t-u)\,\delta'(u^{2}-\epsilon^{2})\\
	&\quad+\frac{\lambda^2}{\pi}\,\chi(t)\int_0^t\!du\;\theta(u)\,\bm{r}(u)\,\chi(t-u)\,X(u)\,\delta''(u^{2}-\epsilon^{2})\,.\notag
\end{align}
Here, as a reminder, the prime of the delta function denotes the derivative with respect to the argument, and thus
\begin{equation*}
	\delta'(u^{2}-\epsilon^{2})=\frac{1}{2u}\frac{d}{du}\delta(u^{2}-\epsilon^2)=\frac{1}{2u}\frac{d}{du}\biggl[\frac{\delta(u-\epsilon)}{2\epsilon}+\frac{\delta(u+\epsilon)}{2\epsilon}\biggr]\,,
\end{equation*}
for $\epsilon>0$. The unit-step function $\theta(u)$ limits the integration to the range $u>0$, so we can drop the $\delta(u+\epsilon)$ term.

The first integral in Eq.~\eqref{E:ybgoe} becomes
\begin{align}\label{E:vbckre}
	-\frac{\lambda^2}{4\pi\epsilon}\,\chi(t) \int_0^t\!du\;\theta(u)\,\frac{\bm{r}(u)\,\chi(t-u)}{u}\,\delta'(u-\epsilon)=\frac{\lambda^2}{4\pi\epsilon}\,\chi(t)\int_0^t\!du\;\frac{d}{du}\biggl[\frac{\bm{r}(u)\,\chi(t-u)}{u}\biggr]\,\delta(u-\epsilon)\,.
\end{align}
We clean up the expressions inside the square brackets, and we have the derivative of the square brackets with respect to $u$ given by
\begin{align}
	\frac{d}{du}\biggl[\frac{\bm{r}(u)\,\chi(t-u)}{u}\biggr]&=-\frac{1}{u^{2}}\,\chi\bm{\epsilon}+\Bigl(\frac{1}{2}\,\ddot{\chi}\bm{\epsilon}-\dot{\chi}\dot{\bm{z}}-\frac{1}{2}\,\chi\ddot{\bm{z}}\Bigr)+u\Bigl(-\frac{1}{3}\,\dddot{\chi}\bm{\epsilon}+\ddot{\chi}\dot{\bm{z}}+\dot{\chi}\ddot{\bm{z}}+\frac{1}{3}\,\chi\dddot{\bm{z}}\Bigr)\,.
\end{align}
The first integral in Eq.~\eqref{E:ybgoe} becomes 
\begin{align}\label{E:oieodqa}
	&\quad-\frac{\lambda^2}{\pi}\,\chi(t)\int_0^t\!du\;\theta(u)\,\bm{r}(u)\,\chi(t-u)\,\delta'(u^{2}-\epsilon^{2})\notag\\
	&=\frac{\lambda^2}{4\pi\epsilon}\,\chi\biggl[-\frac{1}{\epsilon^{2}}\,\chi\bm{\epsilon}+\Bigl(\frac{1}{2}\,\ddot{\chi}\bm{\epsilon}-\dot{\chi}\dot{\bm{z}}-\frac{1}{2}\,\chi\ddot{\bm{z}}\Bigr)+\epsilon\Bigl(-\frac{1}{3}\,\dddot{\chi}\bm{\epsilon}+\ddot{\chi}\dot{\bm{z}}+\dot{\chi}\ddot{\bm{z}}+\frac{1}{3}\,\chi\dddot{\bm{z}}\Bigr)+\mathcal{O}(\epsilon^{2})\biggr]\,.
\end{align}
Before we take the limit $\epsilon\to0$, we take the spherical average to remove any anisotropic contribution. Thus, terms proportional to $\bm{\epsilon}$ will vanish, and we find
\begin{align}\label{E:dijeor}
	&\quad-\frac{\lambda^2}{\pi}\,\chi(t) \int_0^t\!du\;\theta(u)\,\bm{r}(u)\,\chi(t-u)\,\frac{1}{4u\epsilon}\,\delta'(u-\epsilon)\notag\\
    &=-\frac{\lambda^2}{4\pi\epsilon}\Bigl(\chi\dot{\chi}\dot{\bm{z}}+\frac{1}{2}\,\chi^{2}\ddot{\bm{z}}\Bigr)+\frac{\lambda^2}{4\pi}\,\Bigl(\chi\ddot{\chi}\dot{\bm{z}}+\chi\dot{\chi}\ddot{\bm{z}}+\frac{1}{3}\,\chi^{2}\dddot{\bm{z}}\Bigr)+\mathcal{O}(\epsilon)\,.
\end{align}
The second integral in Eq.~\eqref{E:ybgoe} is much more involved to evaluate.

We first re-cast $\delta''(u^{2}-\epsilon^{2})$
\begin{align}
	\delta''(u^{2}-\epsilon^{2})=\frac{1}{2u}\frac{d}{du}\biggl\{\frac{1}{2u}\frac{d}{du}\delta(u^{2}-\epsilon^{2})\biggr\}&=\frac{1}{8u\epsilon}\frac{d}{du}\biggl\{\frac{1}{u}\frac{d}{du}\Bigl[\delta(u-\epsilon)+\delta(u+\epsilon)\Bigr]\biggr\}\notag\\
	&=-\frac{1}{8u^{3}\epsilon}\,\delta'(u-\epsilon)+\frac{1}{8u^{2}\epsilon}\,\delta''(u-\epsilon)\,,
\end{align}
where the contributions associated with $\delta'(u+\epsilon)$ have been dropped. The expansion of $\bm{r}(u)\,\chi(t-u)\,X(u)$ to the order $\mathcal{O}(u^{3})$ is given by 
\begin{align}
	\quad\bm{r}(u)\chi(t-u)X(u)&=u\,K_{1}+u^{2}\,K_{2}+u^{3}\,K_{3}+\mathcal{O}(u^{4})\,,
\end{align}
where 
\begin{align*}
	K_{1}&=2\Bigl(\bm{\epsilon}\cdot\dot{\bm{z}}\Bigr)\,\bm{\epsilon}\,\chi\,,\\
	K_{2}&=2\Bigl(\bm{\epsilon}\cdot\dot{\bm{z}}\Bigr)\Bigl(\dot{\bm{z}}\,\chi-\bm{\epsilon}\,\dot{\chi}\Bigr)+\Bigl(\dot{\bm{z}}^{2}-\bm{\epsilon}\cdot\ddot{\bm{z}}\Bigr)\,\bm{\epsilon}\,\chi\,,\\
	K_{3}&=2\Bigl(\bm{\epsilon}\cdot\dot{\bm{z}}\Bigr)\Bigl(\frac{1}{2}\,\bm{\epsilon}\,\ddot{\chi}-\dot{\bm{z}}\,\dot{\chi}-\frac{1}{2}\,\ddot{\bm{z}}\,\chi\Bigr)+\Bigl(\dot{\bm{z}}^{2}-\bm{\epsilon}\cdot\ddot{\bm{z}}\Bigr)\Bigl(\dot{\bm{z}}\,\chi-\bm{\epsilon}\,\dot{\chi}\Bigr)+\Bigl(\frac{1}{3}\,\bm{\epsilon}\cdot\dddot{\bm{z}}-\dot{\bm{z}}\cdot\ddot{\bm{z}}\Bigr)\,\bm{\epsilon}\,\chi\,.
\end{align*}
Thus, the second integral in Eq.~\eqref{E:ybgoe} becomes 
\begin{align}\label{E:girhisr}
	&\quad\frac{\lambda^2}{\pi}\,\chi(t)\int_0^t\!du\;\theta(u)\,\bm{r}(u)\,\chi(t-u)\,X(u)\,\delta''(u^{2}-\epsilon^{2})\notag\\
	&=\frac{\lambda^2}{8\pi\epsilon}\,\chi(t)\int_0^t\!du\;\biggl\{-\frac{1}{u^{2}}\,K_{1}\,\delta'(u-\epsilon)+\frac{1}{u}\biggl[-K_{2}\,\delta'(u-\epsilon)+K_{1}\,\delta''(u-\epsilon)\biggr]\biggr.\\
    &\qquad\qquad\qquad\qquad+\biggl.\biggl[-K_{3}\,\delta'(u-\epsilon)+K_{2}\,\delta''(u-\epsilon)\biggr]+u\,\biggl[-K_{4}\,\delta'(u-\epsilon)+K_{3}\,\delta''(u-\epsilon)\biggr]+\cdots\biggr\}\,.\notag
\end{align}
We will not include the terms proportional to $\mathcal{O}(u)$ and higher powers in Eq.~\eqref{E:girhisr}, since their surviving contributions, after spherical averaging, are of higher order in velocity and acceleration than the leading nonrelativistic terms considered here.

To evaluate Eq.~\eqref{E:girhisr}, we note that 
\begin{align*}
	\int_0^t\!du\;K_{1}\,\biggl[-\frac{1}{u^{2}}\,\delta'(u-\epsilon)+\frac{1}{u}\,\delta''(u-\epsilon)\biggr]&=0\,,&\int_0^t\!du\;K_{3}\,\biggl[-\delta'(u-\epsilon)+u\,\delta''(u-\epsilon)\biggr]&=0\,,
\end{align*}
so we will focus on the contribution of $K_{2}$, which gives
\begin{align}
	\int_0^t\!du\;K_{2}\,\biggl[-\frac{1}{u}\,\delta'(u-\epsilon)+\delta''(u-\epsilon)\biggr]=-\frac{K_{2}}{\epsilon^{2}}\,.
\end{align}	
Thus, we end up with 
\begin{align}
	\frac{\lambda^2}{\pi}\,\chi(t)\int_0^t\!du\;\theta(u)\,\bm{r}(u)\,\chi(t-u)\,X(u)\,\delta''(u^{2}-\epsilon^{2})&=-\frac{\lambda^2}{8\pi\epsilon^{3}}\,\chi(t)\,K_{2}(t)\,,
\end{align}
where 
\begin{equation*}
	K_{2}=2\bigl(\bm{\epsilon}\cdot\dot{\bm{z}}\bigr)\bigl(\dot{\bm{z}}\,\chi-\bm{\epsilon}\,\dot{\chi}\bigr)+\bigl(\dot{\bm{z}}^{2}-\bm{\epsilon}\cdot\ddot{\bm{z}}\bigr)\,\bm{\epsilon}\,\chi\,.
\end{equation*}
Taking the spherical average over $K_{2}$, we obtain 
\begin{align}
	K_{2}&\overset{\text{s.a.}}{\mapsto}-\frac{2\epsilon^{2}}{3}\,\dot{\bm{z}}\,\dot{\chi}-\frac{\epsilon^{2}}{3}\,\ddot{\bm{z}}\,\chi\,,
\end{align}
and thus we arrive at 
\begin{equation}\label{E:gnhoie}
	\frac{\lambda^2}{\pi}\,\chi(t)\int_0^t\!du\;\theta(u)\,\bm{r}(u)\,\chi(t-u)\,X(u)\,\delta''(u^{2}-\epsilon^{2})=\frac{\lambda^2}{8\pi\epsilon}\,\chi(t)\,\biggl[\frac{2}{3}\,\dot{\bm{z}}\,\dot{\chi}+\frac{1}{3}\,\ddot{\bm{z}}\,\chi\biggr]\,.
\end{equation}

\section{Covariant evaluation of \texorpdfstring{$\phi_{\text{ret}}(z(\tau))$}{phi\_ret(z(tau))}}\label{S:ejheohti}
Since the integral expression in the effective equation of motion Eq.~\eqref{E:eituef} evaluates the self-field directly on the worldline, where it formally diverges, we apply the covariant point-splitting method to regularize it. We define our evaluation point $x^{\mu}$ at proper time $\tau$ by extending it slightly, measured by a small parameter $\epsilon>0$ in the transverse directions of  the worldline,
\begin{equation}
	x^{\mu}=z^{\mu}(\tau)+\epsilon\,n^{\mu}(\tau)\,,
\end{equation}
where $n^{\mu}(\tau)$ is a spacelike unit vector orthogonal to the four-velocity $u^{\mu}(\tau)$, that is, $n_{\mu}n^{\mu}=-1$ and $n_{\mu}u^{\mu}=0$. The retarded time $\tau_{\text{ret}}$ associated with this displaced point is then determined by the light-cone condition, $0=R_{\mu}(\tau_{\text{ret}})R^{\mu}(\tau_{\text{ret}})=\bigl(x-z(\tau_{\text{ret}})\bigr)^{2}$.

Let the proper time delay be $s=\tau-\tau_{\text{ret}}$. We Taylor expand the retarded position $z^{\mu}(\tau_{\text{ret}})$ about the current proper time $\tau$ to obtain
\begin{equation}
    z^{\mu}(\tau_{\text{ret}})=z^{\mu}(\tau)-s\,u^{\mu}(\tau)+\frac{1}{2}\,s^{2}\,a^{\mu}(\tau)-\frac{1}{6}\,s^{3}\,b^{\mu}(\tau)+\frac{1}{24}\,s^{4}\,c^{\mu}(\tau)+\mathcal{O}(s^5)\,,
\end{equation}
where $a^{\mu}=d^{2}z^{\mu}/d\tau^{2}$, $b^{\mu}=d^{3}z^{\mu}/d\tau^{3}$, and $c^{\mu}=d^{4}z^{\mu}/d\tau^{4}$, respectively. The separation vector then becomes
\begin{align*}
    R^{\mu}(\tau_{\text{ret}})&=x^{\mu}-z^{\mu}(\tau_{\text{ret}})=\epsilon\,n^{\mu}(\tau)+s\,u^{\mu}(\tau)-\frac{1}{2}\,s^{2}\,a^{\mu}(\tau)+\frac{1}{6}\,s^{3}\,b^{\mu}(\tau)-\frac{1}{24}\,s^{4}\,c^{\mu}(\tau)+\mathcal{O}(s^5)\,.		
\end{align*}
Applying the light-cone condition $R^{\mu}R_{\mu}=0$ and utilizing the kinematic constraints $u^{2}=1$, $n^{2}=-1$, $n\cdot u=0$, and $u\cdot a=0$, which implies $u\cdot b=-a^2$, we find
\begin{align}\label{E:toeirf}
    R^{2}=0&=-\epsilon^{2}+s^{2}\,\bigl(1-\epsilon\,n\cdot a\bigr)+\frac{1}{3}\,s^{3}\,\epsilon\,n\cdot b-\frac{1}{12}\,s^{4}\bigl(\epsilon\,n\cdot c+a^{2}\bigr)+\mathcal{O}(s^5)\,,	
\end{align}
where we have adopted the shorthand notation for Lorentz scalars $A \cdot B=A_{\mu}B^{\mu}$. To solve for $s$ in terms of $\epsilon$, we assume a power series ansatz
\begin{equation*}
    s=c_{1}\,\epsilon+c_{2}\,\epsilon^{2}+c_{3}\,\epsilon^{3}+\cdots,,
\end{equation*}
and substitute this back into Eq.~\eqref{E:toeirf}. Equating coefficients of the same order in $\epsilon$ yields
\begin{align*}
&\mathcal{O}(\epsilon^{2}):&c_{1}&=1\,,\\
&\mathcal{O}(\epsilon^{3}):&c_{2}&=\frac{1}{2}\,n\cdot a\,,\\
&\mathcal{O}(\epsilon^{4}):&c_{3}&=-\frac{1}{6}\,n\cdot b+\frac{3}{8}\,(n\cdot a)^{2}+\frac{1}{24}\,a^{2}\,.	
\end{align*}
Thus, the time delay expansion is
\begin{equation}\label{E:ngns}
    s=\epsilon+\frac{1}{2}\,\bigl(n\cdot a\bigr)\,\epsilon^{2}+\biggl[-\frac{1}{6}\,n\cdot b+\frac{3}{8}(n\cdot a)^{2}+\frac{1}{24}\,a^{2}\biggr]\,\epsilon^{3}+\mathcal{O}(\epsilon^4)\,.
\end{equation}
Next, we expand the retarded invariant distance $r(\tau_{\text{ret}})=R_{\mu}(\tau_{\text{ret}})u^{\mu}(\tau_{\text{ret}})$ in powers of $\epsilon$. Using the expansion for the retarded four-velocity,
\begin{equation*}
    u^{\mu}(\tau_{\text{ret}})=u^{\mu}(\tau)-s\,a^{\mu}(\tau)+\frac{1}{2}\,s^{2}\,b^{\mu}(\tau)-\frac{1}{6}\,s^{3}\,c^{\mu}(\tau)+\mathcal{O}(s^4)\,,
\end{equation*}
we evaluate the inner product
\begin{align}
    r(\tau_{\text{ret}})=R_{\mu}(\tau_{\text{ret}})u^{\mu}(\tau_{\text{ret}})&=\biggl[\epsilon\,n_{\mu}+s\,u_{\mu}-\frac{1}{2}s^{2}a_{\mu}+\frac{1}{6}s^{3}b_{\mu}+\mathcal{O}(s^4)\biggr]\notag\\
                &\qquad\qquad\qquad\cdot\biggl[u^{\mu}-s\,a^{\mu}+\frac{1}{2}s^{2}b^{\mu}-\frac{1}{6}s^{3}c^{\mu}+\mathcal{O}(s^4)\biggr]\notag\\
                &=\epsilon-\frac{1}{2}\bigl(n\cdot a\bigr)\epsilon^{2}+\mathcal{O}(\epsilon^{3})\,,
\end{align}
where we have substituted Eq.~\eqref{E:ngns} and truncated at the appropriate order.

Finally, we expand the internal state $\chi(\tau_{\text{ret}})$ around the current proper time $\tau$. Using the time delay expansion $s=\tau-\tau_{\text{ret}}$, we find
\begin{equation}
	\chi(\tau_{\text{ret}})=\chi(\tau-s)=\chi(\tau)-s\,\dot{\chi}(\tau)+\mathcal{O}(s^{2})\,.
\end{equation}
Substituting this and the expansion for $r(\tau_{\text{ret}})$ into the Li\'enard-Wiechert potential, the retarded field $\phi_{\text{ret}}(x)$ can be expressed as a series in the point-splitting parameter $\epsilon$,
\begin{align*}
	\phi_{\text{ret}}(z(\tau))=\frac{\lambda}{4\pi}\frac{\chi(\tau_{\text{ret}})}{r(\tau_{\text{ret}})}=\frac{\lambda}{4\pi}\frac{\chi(\tau)-\dot{\chi}(\tau)\,\epsilon+\mathcal{O}(\epsilon^{2})}{\epsilon-\dfrac{1}{2}\,\bigl(n\cdot a\bigr)\,\epsilon^{2}+\mathcal{O}(\epsilon^{3})}=\frac{\lambda}{4\pi}\biggl[\frac{\chi(\tau)}{\epsilon}-\dot{\chi}(\tau)+\frac{1}{2}\,\chi(\tau)\,\bigl(n\cdot a\bigr)+\mathcal{O}(\epsilon)\biggr]\,.
\end{align*}	
If the internal degree of freedom acts as an idealized scalar monopole, the point-splitting regularization must be applied isotropically. We therefore average over all possible spatial directions $n^{\mu}$ in the proper rest frame of the detector. Since the spatial distribution is spherically symmetric, the angular average of the unit normal vector vanishes, $\langle n^{\mu}\rangle=0$. Taking the limit as $\epsilon\to0$ for the finite terms, we arrive at the regularized retarded field on the worldline,
\begin{equation}
	\phi_{\text{ret}}(z(\tau))=\frac{\lambda}{4\pi}\frac{\chi(\tau)}{\epsilon}-\frac{\lambda}{4\pi}\,\frac{d}{d\tau}\chi(\tau)\,.
\end{equation}

\section{Variations of the actions}\label{S:fbgsfs}
Here we will compute the variation of the detector's action
\begin{align}
    S_{\text{detector}}=\int\!d\varsigma\;\sqrt{z'^2}\,\biggl\{-\biggl[M+V(z)\biggr]+\biggl[\frac{m}{2}\,\biggl(\frac{\chi'}{\sqrt{z'^2}}\biggr)^2-\frac{m\omega^2}{2}\,\chi^2\biggr]+\lambda\,\chi\,\phi(z)\biggr\}\,.
\end{align}
with respect to $z^{\mu}$.

The variation of the measure $\sqrt{z'^2}$ with respect to $z^{\mu}$ is
\begin{equation}
	\delta\sqrt{z'^2}=\frac{z'_{\mu}}{\sqrt{z'^2}}\,\delta z'^{\mu}=\frac{z'_{\mu}}{\sqrt{z'^2}}\,\frac{d}{d\varsigma}\delta z^{\mu}\,.
\end{equation}
Applying this, the variation of the mechanical degree of freedom yields
\begin{align}\label{E:ngsr}
    \delta S_{\text{mdf}}&=-\int\!d\varsigma\;\biggl\{M\,\delta\sqrt{z'^2}+V(z)\,\delta\sqrt{z'^2}+\sqrt{z'^2}\,\partial_{\mu}V(z)\,\delta z^{\mu}\biggr\}\notag\\
    &=\int\!d\varsigma\;\biggl\{-\Bigl[M+V(z)\Bigr]\,\frac{z'_{\mu}}{\sqrt{z'^2}}\,\frac{d}{d\varsigma}\delta z^{\mu}-\sqrt{z'^2}\,\partial_{\mu}V(z)\,\delta z^{\mu}\biggr\}\notag\\
    &=\int\!d\varsigma\;\biggl\{M\,\frac{d}{d\varsigma}\biggl(\frac{z'_{\mu}}{\sqrt{z'^2}}\biggr)+\frac{d}{d\varsigma}\biggl[V(z)\,\frac{z'_{\mu}}{\sqrt{z'^2}}\biggr]-\sqrt{z'^2}\,\partial_{\mu}V(z)\biggr\}\,\delta z^{\mu}\notag\\
    &=\int\!d\tau\;\biggl\{M\,\ddot{z}_{\mu}+\frac{d}{d\tau}\biggl[V(z)\,\dot{z}_{\mu}\biggr]-\partial_{\mu}V(z)\biggr\}\,\delta z^{\mu}\,,
\end{align}
where we have restored the overdot notation for the derivative with respect to the proper time and utilized the identities
\begin{align}
    \dot{z}&=\frac{dz^{\mu}}{d\tau}=\frac{z'^{\mu}}{\sqrt{z'^2}}\,,&\ddot{z}_{\mu}&=\frac{d^2z^{\mu}}{d\tau^2}=\frac{1}{\sqrt{z'^2}}\frac{d}{d\varsigma}\biggl(\frac{z'_{\mu}}{\sqrt{z'^2}}\biggr)\,.
\end{align}
Similarly, the variation of the internal degree of freedom is
\begin{align}\label{E:ndfjet}
    \delta S_{\text{idf}}&=\int\!d\varsigma\;\biggl\{\biggl[\frac{m}{2}\,\biggl(\frac{\chi'}{\sqrt{z'^2}}\biggr)^2-\frac{m\omega^2}{2}\,\chi^2\biggr]\,\frac{z'_{\mu}}{\sqrt{z'^2}}\,\frac{d}{d\varsigma}\delta z^{\mu}-m\,\biggl(\frac{\chi'}{\sqrt{z'^2}}\biggr)^2\,\frac{z'_{\mu}}{\sqrt{z'^2}}\,\frac{d}{d\varsigma}\delta z^{\mu}\biggr\}\notag\\
    &=\int\!d\varsigma\;\frac{d}{d\varsigma}\biggl\{\biggl[\frac{m}{2}\,\biggl(\frac{\chi'}{\sqrt{z'^2}}\biggr)^2+\frac{m\omega^2}{2}\,\chi^2\biggr]\,\frac{z'_{\mu}}{\sqrt{z'^2}}\biggr\}\,\delta z^{\mu}\notag\\
    &=\int\!d\tau\;\frac{d}{d\tau}\biggl\{\biggl[\frac{m}{2}\,\dot{\chi}^2+\frac{m\omega^2}{2}\,\chi^2\biggr]\,\dot{z}_{\mu}\biggr\}\,\delta z^{\mu}
\end{align}
Finally, we compute the variation of the interaction action
\begin{equation}
    S_{\text{int}}=\int\!d\varsigma\;\sqrt{z'^2}\,\biggl\{\lambda\,\chi\,\phi(z)\biggr\}\,,
\end{equation}
which takes the form
\begin{align}\label{E:jetujet}
	\delta S_{\text{int}}&=\lambda\int\!d\varsigma\;\biggl\{\chi\,\partial_{\mu}\phi(z)\,\sqrt{z'^{2}}\,\delta z^{\mu}+\chi\,\phi(z)\,\frac{z'_{\mu}}{\sqrt{z'^{2}}}\,\frac{d}{d\varsigma}\delta z^{\mu}\biggr\}\notag\\
    &=\lambda\int\!d\varsigma\;\sqrt{z'^{2}}\,\biggl\{\chi\,\partial_{\mu}\phi(z)-\frac{1}{\sqrt{z'^{2}}}\frac{d}{d\varsigma}\biggl[\chi\,\phi(z)\,\frac{z'_{\mu}}{\sqrt{z'^{2}}}\biggr]\biggr\}\,\delta z^{\mu}\notag\\
    &=\lambda\int\!d\tau\;\biggl\{\chi\,\partial_{\mu}\phi(z)-\frac{d}{d\tau}\biggl[\chi\,\phi(z)\,\dot{z}_{\mu}\biggr]\biggr\}\,\delta z^{\mu}\,.
\end{align}

\section{Pauli-Villars regularization}\label{S:eothgbd}
The core idea of Pauli-Villars regularization is to introduce an ancillary scalar field with a large fictitious mass $\Lambda$. The Green's function for this massive field satisfies
\begin{equation}\label{E:dngoes}
	\bigl(\Box+\Lambda^2\bigr)G^{(\phi)}_{\Lambda}(x,x')=\delta^{(4)}(x-x')\,,
\end{equation}
in contrast to the original massless Green's function $G^{(\phi)}_{0}$ which satisfies $\Box G^{(\phi)}_{0}(x,x')=\delta^{(4)}(x-x')$. The regularized Green's function is then defined as the difference between the two,	
\begin{equation}
	G^{(\phi)}_{\text{reg}}(x,x')=G^{(\phi)}_{0}(x,x')-G^{(\phi)}_{\Lambda}(x,x')\,.
\end{equation}
This ancillary contribution $G^{(\phi)}_{\Lambda}$ is mathematically designed to suppress the high-frequency modes or equivalently, the short-distance singularities of the original Green's function $G^{(\phi)}_{0}$. As a result, the regularized Green's function $G^{(\phi)}_{\text{reg}}$ becomes a finite, fully Lorentz-covariant function of the spacetime coordinates everywhere, including at the detector's location.

The regularized field $\phi_{\text{reg}}$ is obtained by convolving the regularized Green's function with the source,
\begin{equation}
	\phi_{\text{reg}}(x)=\int\!d^4x'\;G^{(\phi)}_{\text{reg}}(x,x')\,J(x')\,.
\end{equation}
Since $G^{(\phi)}_{\text{reg}}$ is finite, $\phi_{\text{reg}}$ is also finite across the detector's trajectory, allowing us to safely evaluate the field. Its gradient, however, must still be evaluated using a consistent distributional coincidence prescription.

The explicit form of the regularized retarded Green's function is given by
\begin{equation}\label{E:noqoqex}
	G^{(\phi)}_{\text{reg}}(x,x')=\frac{\Lambda}{4\pi\sigma}\,\theta(\Delta t)\,\theta(\sigma^{2})\,J_{1}(\Lambda\sigma)\,.
\end{equation}
where $\sigma^{\mu}=x^{\mu}-x'^{\mu}$ is the coordinate separation, and $\sigma^{2}=\sigma_{\mu}\sigma^{\mu}=(\sigma^0)^{2}-\vert{}\bm{\sigma}\vert{}^{2}$ is the squared invariant spacetime interval. If the source carries charge $\mathfrak{d}(\tau)$ moving along a worldline $z(\tau)$ parameterized by proper time $\tau$, its source density is
\begin{equation}
	J(x)=\int\!d\tau\;\mathfrak{d}(\tau)\,\delta^{(4)}(x^{\mu}-z^{\mu}(\tau))\,.
\end{equation}
Integrating this against the Green's function yields the regularized field
\begin{equation}
	\phi_{\text{reg}}(x)=\int\!d^4x'\;G^{(\phi)}_{\text{reg}}(x,x')\,J(x')=\int\!d\tau'\;\mathfrak{d}(\tau')\,G^{(\phi)}_{\text{reg}}(x,z(\tau'))\,.
\end{equation}	
Next, we evaluate the four-gradient of the regularized field, $\partial^{\mu}\phi_{\text{reg}}$.

Applying the chain rule, we have the four-gradient of the regularized field given by 
\begin{align}\label{E:xeorer}
	\partial^{\mu}\phi_{\text{reg}}(x)=\int\!d\tau'\;\mathfrak{d}(\tau')\,\biggl[\frac{\partial}{\partial\sigma'^{2}}G^{(\phi)}_{\text{reg}}(\sigma'^{2})\biggr]\,\partial^{\mu}\sigma'^{2}\,.
\end{align}
Hereafter, we introduce the shorthand notations $\sigma'=\sigma^{2}(\tau')=[x_{\nu}-z_{\nu}(\tau')][x^{\nu}-z^{\nu}(\tau')]$, and $\partial_{\sigma'^2}G^{(\phi)}_{\text{reg}}(\sigma^{2})=\partial G^{(\phi)}_{\text{reg}}(\sigma'^{2})/\partial\sigma'^{2}$. The derivative of the squared interval with respect to the observation point coordinate $x_{\mu}$ is given by $\partial^{\mu}\sigma'^{2}=2\bigl[x^{\mu}-z^{\mu}(\tau')\bigr]$. We define the separation vector pointing from the source to the observation point as $\sigma'^{\mu}(\tau)=x^{\mu}-z^{\mu}(\tau')$. Thus, we arrive at 
\begin{equation}\label{E:fhetoe}
	\partial^{\mu}\phi_{\text{reg}}(x)=2\int\!d\tau'\;\mathfrak{d}(\tau')\,\sigma^{\mu}(\tau')\,\partial_{\sigma'^2}G^{(\phi)}_{\text{reg}}(\sigma'^{2})\,.
\end{equation}
Using the explicit definition of $G^{(\phi)}_{\text{reg}}$, we compute the derivative with respect to $\sigma'^2$,
\begin{align}\label{E:ghoeuets}
	\partial_{\sigma'^2}G^{(\phi)}_{\text{reg}}(\sigma'^{2})=\frac{\Lambda}{4\pi}\,\theta(\sigma'^{0})\frac{\partial}{\partial\sigma'^{2}}\biggl[\theta(\sigma'^{2})\,\frac{J_{1}(\Lambda\sqrt{\sigma'^{2}})}{\sqrt{\sigma'^{2}}}\biggr]=\frac{\Lambda^{2}}{8\pi}\,\theta(\sigma'^{0})\,\biggl\{\delta(\sigma'^{2})-\theta(\sigma'^{2})\,\frac{J_{2}(\Lambda\sigma')}{\sigma'^{2}}\biggr\}\,.
\end{align}
Substituting this back into the integral Eq.~\eqref{E:fhetoe}, we find
\begin{align}\label{E:dljndgoet}
	\partial^{\mu}\phi_{\text{reg}}(x)=\frac{\Lambda^{2}}{4\pi}\int_{-\infty}^{\infty}\!d\tau'\;\theta(\sigma'^{0})\,\mathfrak{d}(\tau')\,\sigma^{\mu}(\tau')\,\biggl\{\delta(\sigma'^{2})-\theta(\sigma'^{2})\,\frac{J_{2}(\Lambda\,\sigma')}{\sigma'^{2}}\biggr\}\,.
\end{align}
Note that there is an additional contribution in $\partial^{\mu}G^{(\phi)}_{\text{reg}}(\sigma'^{2})$ due to the gradient of $\theta(\sigma'^{0})$, 
\begin{align}
	\partial^{\mu}\theta(\sigma'^{0})=\delta^{\mu0}\,\delta(\sigma'^{0})\,.
\end{align}
This makes the factor $\theta(\sigma'^{2})=\theta(-\lvert\bm{\sigma}'\rvert^2)$ vanish for all noncoincident spatial separations, because $-\lvert\bm{\sigma}'\rvert^2<0$. Thus, it does not contribute to the gradient of the regularized field, and we can safely ignore this contribution.

Owing to the causal step function $\theta(\sigma'^0)$, the integrand vanishes for any proper time $\tau'$ in the future of the observation point $x$. We can therefore absorb this step function by truncating the upper limit of the integral at the retarded proper time $\tau_{\text{ret}}$,
\begin{equation}\label{E:mvntiue}
	\partial^{\mu}\phi_{\text{reg}}(x)=\frac{\Lambda^{2}}{4\pi}\int_{-\infty}^{\tau_{\text{ret}}}\!d\tau'\;\mathfrak{d}(\tau')\,\sigma^{\mu}(\tau')\,\biggl\{\delta(\sigma'^{2})-\theta(\sigma'^{2})\,\frac{J_{2}(\Lambda\,\sigma')}{\sigma'^{2}}\biggr\}\,.
\end{equation} 
In the end, when we evaluate this gradient on the worldline of the detector to find the self-force, i.e., setting the observation point $x^{\mu} \to z^{\mu}(\tau)$, the retarded time $\tau_{\text{ret}}$ associated with the radiation field that can reach $x^{\mu}$ simply converges to the detector's present proper time $\tau$.

Let us first calculate the contribution of the $\delta(\sigma'^{2})$ term in Eq.~\eqref{E:mvntiue}, evaluated directly on the worldline, that is, $x^{\mu}=z^{\mu}(\tau)$. Defining the proper time delay as $s=\tau-\tau'$, we can rewrite the integral as
\begin{equation}
	\frac{\Lambda^{2}}{4\pi}\int_{-\infty}^{\tau}\!d\tau'\;\mathfrak{d}(\tau')\,\sigma^{\mu}(\tau')\,\delta(\sigma'^{2})=\frac{\Lambda^{2}}{4\pi}\int_{0}^{\infty}\!ds\;\mathfrak{d}(\tau-s)\,\sigma^{\mu}(s;\tau)\,\delta(\sigma^{2}(s))\,.
\end{equation}
Now we expand the $s$-dependent expressions around the present moment $s=0$, i.e., the current proper time $\tau$,
\begin{align}
	\mathfrak{d}(\tau-s)&=\mathfrak{d}-s\,\dot{\mathfrak{d}}+\frac{s^{2}}{2}\,\ddot{\mathfrak{d}}+\mathcal{O}(s^{3})\,,\\
	\sigma^{\mu}(s)=z^{\mu}(\tau)-z^{\mu}(\tau-s)&=s\,u^{\mu}-\frac{s^{2}}{2}\,a^{\mu}+\mathcal{O}(s^{3})\,.
\end{align}
where $u^{\mu}$ is the four-velocity. Applying the constraint $u_{\mu}a^{\mu}=0$, the squared invariant interval expands as $\sigma^{2}=s^{2}+\mathcal{O}(s^{4})$. This implies that the delta function can be approximated near the lightcone as 
\begin{equation}
	\delta(\sigma^{2})\approx\delta(s^{2})=\frac{1}{2s}\,\delta(s)\,.
\end{equation}
Multiplying the terms together, we find
\begin{align}
	\mathfrak{d}(\tau-s)\,\sigma^{\mu}(s;\tau)\,\delta(\sigma^{2}(s))&=\biggl(\mathfrak{d}-s\,\dot{\mathfrak{d}}+\frac{s^{2}}{2}\,\ddot{\mathfrak{d}}+\cdots\biggr)\biggl(s\,u^{\mu}-\frac{s^{2}}{2}\,a^{\mu}+\cdots\biggr)\,\frac{\delta(s)}{2s}=\frac{\mathfrak{d}\,u^{\mu}}{2}\,\delta(s)\,.
\end{align}
Integrating this over $s\in [0,\infty)$ yields a factor of $1/2$ from the boundary of the delta function,
\begin{align}\label{E:vboeuessq}
	\frac{\Lambda^{2}}{4\pi}\int_{-\infty}^{\tau}\!d\tau'\;\mathfrak{d}(\tau')\,\sigma^{\mu}(\tau')\,\delta(\sigma'^{2})=\frac{\Lambda^{2}}{4\pi}\int_{0}^{\infty}\!ds\;\frac{\mathfrak{d}\,u^{\mu}}{2}\,\delta(s)=\frac{\Lambda^{2}}{16\pi}\,\mathfrak{d}(\tau)\,u^{\mu}(\tau)\,.
\end{align}
Note that this result depends only on the instantaneous state of the detector and not on its time derivatives, as the field is sampled exactly on the light cone, $s=0$.

While this contribution is proportional to $\Lambda^{2}$ and represents the most severe divergence in the $\Lambda\to\infty$ limit, it is also proportional to the four-velocity $u^{\mu}$. Consequently, it will not affect the mechanical (external) dynamics. When substituted into the external equation of motion, the transverse projection projector annihilates it completely since $P_{\mu\nu}u^{\nu} = 0$. Thus, this $\mathcal{O}(\Lambda^{2})$ divergence exerts absolutely no kinematic force on the detector and drops out of the observable dynamics entirely, without even requiring mass renormalization.

The second contribution to the gradient in Eq.~\eqref{E:mvntiue} is given by
\begin{align}\label{E:bvier}
	-\frac{\Lambda^{2}}{4\pi}\int_{-\infty}^{\tau}\!d\tau'\;\mathfrak{d}(\tau')\,\sigma^{\mu}(\tau')\,\theta(\sigma'^{2})\,\frac{J_{2}(\Lambda\,\sigma')}{\sigma'^{2}}\,,
\end{align}
where $\sigma^{\mu}=z^{\mu}(\tau)-z^{\mu}(\tau')$. Since the macroscopic motion of the detector is timelike, the invariant interval satisfies $\sigma^{2}(\tau')>0$ for all past times $\tau'<\tau$. Thus, $\theta(\sigma'^{2})=1$ within the domain of integration, and this step function can be safely dropped. Equation~\eqref{E:bvier} then reduces to
\begin{align}\label{E:birts}
	-\frac{\Lambda^{2}}{4\pi}\int_{-\infty}^{\tau}\!d\tau'\;\mathfrak{d}(\tau')\,\sigma^{\mu}(\tau')\,\frac{J_{2}(\Lambda\,\sigma')}{\sigma'^{2}}\,.
\end{align}
As before, we expand the integrand in terms of the proper time delay $s=\tau-\tau'$, and have
\begin{align}
	\sigma^{\mu}(s)&=s\,u^{\mu}-\frac{s^{2}}{2}\,a^{\mu}+\frac{s^{3}}{6}\,b^{\mu}-\frac{s^{4}}{24}\,c^{\mu}+\mathcal{O}(s^{4})\,,\\
	\mathfrak{d}(\tau-s)&=\mathfrak{d}(\tau)-s\,\dot{\mathfrak{d}}(\tau)+\frac{s^{2}}{2}\,\ddot{\mathfrak{d}}(\tau)-\frac{s^{3}}{6}\,\dddot{\mathfrak{d}}(\tau)+\mathcal{O}(s^{4})\,,\\
	\sigma^{2}(s)&=s^{2}+s^{3}\bigl(u\cdot a\bigr)-s^{4}\biggl(\frac{1}{4}\,a^{2}+\frac{1}{3}\,u\cdot b\biggr)+\mathcal{O}(s^{5})=s^{2}-\frac{a^{2}}{12}\,s^{4}+\frac{a\cdot b}{12}\,s^{5}+\mathcal{O}(s^{6})\,,\\
	\sigma(s)&=s\,\biggl(1-\frac{a^{2}1}{12}\,s^{2}+\frac{a\cdot b}{12}\,s^{3}+\cdots\biggr)^{\frac{1}{2}}=s-\frac{a^{2}}{24}\,s^{3}+\frac{a\cdot b}{24}\,s^{4}+\mathcal{O}(s^{5})\,.
\end{align}
Then the ratio $\sigma^{\mu}/\sigma^{2}$ expands as
\begin{align}
	\frac{\sigma^{\mu}}{\sigma^{2}}&=\frac{u^{\mu}}{s}-\frac{a^{\mu}}{2}+\biggl[\frac{1}{12}\,a^{2}\,u^{\mu}+\frac{1}{6}\,b^{\mu}\biggr]\,s+\biggl[-\frac{1}{24}\,a^{2}\,a^{\mu}-\frac{1}{12}\,\bigl(a\cdot b\bigr)\,u^{\mu}-\frac{1}{24}\,c^{\mu}\biggr]\,s^{2}+\mathcal{O}(s^{3})\,.
\end{align}
We do not want to fully expand the Bessel function $J_{2}(\Lambda\sigma)$ with respect to $s$ because we must retain its asymptotic behavior at large arguments. Instead, we expand its argument 
\begin{align}
	J_{2}(\Lambda\sigma)\simeq J_{2}(\Lambda s)-\frac{\Lambda a^{2}}{24}\,s^{3}\,J'_{2}(\Lambda s)+\cdots\,.
\end{align}
This expansion might naively seem problematic in the limit $\Lambda\to\infty$. However, if we change the integration variable to $y=\Lambda s$, then the expansion becomes valid,
\begin{align}
	\Lambda\,\sigma=\Lambda\,s-\frac{\Lambda\,a^{2}}{24}\,s^{3}+\frac{\Lambda\,a\cdot b}{24}\,s^{4}+\cdots=y-\frac{a^{2}}{24\Lambda^{2}}\,y^{3}+\frac{a\cdot b}{24\Lambda^{3}}\,y^{4}+\cdots\,.
\end{align}
Thus, the corrections are highly suppressed by powers of $\Lambda^{-2}$ as $\Lambda\to\infty$.  Furthermore, the integral of the product of a polynomial and the Bessel $J_{n}$ is well defined. Rewriting Eq.~\eqref{E:birts} in terms of $y$, where $ds=dy/\Lambda$, we obtain
\begin{align}
	&\quad-\frac{\Lambda^{2}}{4\pi}\int_{0}^{\infty}\!ds\;\mathfrak{d}(\tau-s)\,\sigma^{\mu}(s;\tau)\,\frac{J_{2}(\Lambda\,\sigma(s))}{\sigma'^{2}(s)}\notag\\
	&=\frac{1}{4\pi}\int_{0}^{\infty}\!dy\;\biggl\{-\mathfrak{d}\,u^{\mu}\,\frac{J_{2}(y)}{y}\,\Lambda^{2}+\biggl[\frac{\mathfrak{d}}{2}\,a^{\mu}\,J_{2}(y)+\dot{\mathfrak{d}}\,u^{\mu}\,J_{2}(y)\biggr]\,\Lambda\biggr.\\
	&\qquad\qquad\qquad\qquad+\biggl.\biggl[-\frac{\mathfrak{d}\,a^{2}}{24}\,u^{\mu}\,y^{2}J_{3}(y)-\frac{\mathfrak{d}}{6}\,b^{\mu}\,y\,J_{2}(y)-\frac{\dot{\mathfrak{d}}}{2}\,a^{\mu}\,y\,J_{2}(y)-\frac{\ddot{\mathfrak{d}}}{2}\,u^{\mu}\,y\,J_{2}(y)\biggr]+\mathcal{O}(\Lambda^{-1})\biggr\}\,,\notag
\end{align}
where we have applied standard Bessel derivative identities to express $J'_2(y)$ terms in terms of $J_3(y)$. Evaluating this requires the following standard integral identities
\begin{align}
	\int_{0}^{\infty}\!dx\;\frac{J_{2}(y)}{y}&=\frac{1}{2}\,,&\int_{0}^{\infty}\!dx\;J_{2}(y)&=1\,.
\end{align}
However, the proper time expansions have generated moments of the Bessel function, specifically,
\begin{align*}
	&\int_{0}^{\infty}\!dx\;y\,J_{2}(y)\,,&&\int_{0}^{\infty}\!dx\;y^{2}J_{3}(y)\,,
\end{align*}
which are formally divergent integrals. Thus, even after utilizing the regularized retarded Green's function, we encounter conditionally divergent integrals when evaluating the radiation reaction. This raises a subtle question of mathematical regularization: if we introduce a convergence factor to define these integrals, does the finite result depend on the choice of the cutoff?

Let us examine the source of the divergence. In the large $y$ limit, the asymptotic behaviors of the Bessel functions are 
\begin{align}
	J_{2}(y)&\simeq -\sqrt{\frac{2}{\pi}}\,y^{-\frac{1}{2}}\,\cos(y-\frac{\pi}{4})+\sqrt{\frac{2}{\pi}}\frac{15}{8}\,y^{-\frac{3}{2}}\,\sin(y)+\mathcal{O}(y^{-\frac{5}{2}})\,,\\
	J_{3}(y)&\simeq +\sqrt{\frac{2}{\pi}}\,y^{-\frac{1}{2}}\,\cos(y+\frac{\pi}{4})-\sqrt{\frac{2}{\pi}}\frac{35}{8}\,y^{-\frac{3}{2}}\,\sin(y+\frac{\pi}{4})+\mathcal{O}(y^{-\frac{5}{2}})\,.
\end{align}
The leading terms of $y\,J_{2}(y)$ and $y^{2}\,J_{3}(y)$ grow as $y^{1/2}$ and $y^{3/2}$, respectively, causing the integrals to oscillate infinitely with growing amplitude. To define these limits, we insert an exponential convergence factor $e^{-\epsilon y}$ into the integrands, evaluated in the limit as $\epsilon\to0$
\begin{align}
	\lim_{\epsilon\to0}\int_{0}^{\infty}\!dy\;e^{-\epsilon y}\,y\,J_{2}(y)&=2\,,&\lim_{\epsilon\to0}\int_{0}^{\infty}\!dy\;e^{-\epsilon y}\,y^2\,J_{3}(y)&=8\,,
\end{align}
These two integrals do not have any contributions that blow up as $\epsilon\to0$, so we can safely take the limit $\epsilon\to0$ and accept the finite results. Applying these evaluated integrals yields
\begin{align}
	\frac{\Lambda^{2}}{4\pi}\int_{0}^{\infty}\!ds\;\mathfrak{d}(\tau-s)\,\sigma^{\mu}(s;\tau)\,\frac{J_{2}(\Lambda\,\sigma(s))}{\sigma'^{2}(s)}&=-\frac{1}{8\pi}\,\mathfrak{d}\,u^{\mu}\,\Lambda^{2}+\biggl[\frac{1}{8\pi}\,\mathfrak{d}\,a^{\mu}+\frac{1}{4\pi}\,\dot{\mathfrak{d}}\,u^{\mu}\biggr]\,\Lambda\notag\\
    &\qquad\qquad+\biggl[-\frac{1}{12\pi}\,\mathfrak{d}\,a^{2}\,u^{\mu}-\frac{1}{12\pi}\,\mathfrak{d}\,b^{\mu}-\frac{1}{4\pi}\,\dot{\mathfrak{d}}\,a^{\mu}-\frac{1}{4\pi}\,\ddot{\mathfrak{d}}\,u^{\mu}\biggr]\notag\\
    &\qquad\qquad\qquad\qquad+\mathcal{O}(\Lambda^{-1})\,.
\end{align}
Finally, combining this with the contribution from Eq.~\eqref{E:vboeuessq}, we arrive at the total self-field gradient
\begin{align}\label{E:dfdf}
	\partial^{\mu}\phi_{\text{reg}}(x)&=-\frac{1}{16\pi}\,\mathfrak{d}\,u^{\mu}\,\Lambda^{2}+\biggl[\frac{1}{8\pi}\,\mathfrak{d}\,a^{\mu}+\frac{1}{4\pi}\,\dot{\mathfrak{d}}\,u^{\mu}\biggr]\,\Lambda\notag\\
    &\qquad\qquad\qquad+\biggl[-\frac{1}{12\pi}\,\mathfrak{d}\,a^{2}\,u^{\mu}-\frac{1}{12\pi}\,\mathfrak{d}\,b^{\mu}-\frac{1}{4\pi}\,\dot{\mathfrak{d}}\,a^{\mu}-\frac{1}{4\pi}\,\ddot{\mathfrak{d}}\,u^{\mu}\biggr]+\mathcal{O}(\Lambda^{-1})\,.
\end{align}

\newpage
\bibliography{refs23}
\end{document}